\documentclass[%
 reprint,
 amsmath,amssymb,
 aps,
]{revtex4-2}

\usepackage{graphicx}
\usepackage{dcolumn}
\usepackage{bm}
\usepackage{multirow} 

\begin{document}
\def\slopep{\Delta\mathrm{HR}/ \Delta\mathrm{age}}
\def\magGyr{\,\mathrm{mag/Gyr}}
\title{\boldmath Revisiting progenitor-age dependence of type Ia supernova luminosity standardization process}

\author{Junchao Wang}
 \affiliation{School of Physics and Astronomy, Sun Yat-sen University,\\
Zhuhai, 519082, P.R.China}
 \affiliation{CSST Science Center for the Guangdong-Hongkong-Macau Greater Bay Area, Sun Yat-sen University,\\
Zhuhai, 519082, P.R.China}
\author{Zhiqi Huang}
 \email{huangzhq25@mail.sysu.edu.cn}
 \affiliation{School of Physics and Astronomy, Sun Yat-sen University,\\
Zhuhai, 519082, P.R.China}
 \affiliation{CSST Science Center for the Guangdong-Hongkong-Macau Greater Bay Area, Sun Yat-sen University,\\
Zhuhai, 519082, P.R.China}
\author{Lu Huang}
 \affiliation{School of Physics and Astronomy, Sun Yat-sen University,\\
Zhuhai, 519082, P.R.China}
 \affiliation{CSST Science Center for the Guangdong-Hongkong-Macau Greater Bay Area, Sun Yat-sen University,\\
Zhuhai, 519082, P.R.China}

\begin{abstract}
Much of the research in supernova cosmology is based on an assumption that the peak luminosity of type Ia supernovae (SNe Ia), after a standardization process, is independent of the galactic environment. A series of recent studies suggested that there is a significant correlation between the standardized luminosity and the progenitor age of SNe Ia. The correlation found in the most recent work by Lee et al. is strong enough to explain the extra dimming of distant SNe Ia and therefore casts doubts on the direct evidence of cosmic acceleration. The present work incorporates the uncertainties of progenitor ages, which were ignored in Lee et al., into a fully Bayesian inference framework. We find a weaker dependence of supernova standardized luminosity on the progenitor age,  but the detection of correlation remains significant (3.5$\sigma$). Assuming that such correlation can be extended to high redshift and applying it to the Pantheon SN Ia data set, we confirm that when the Hubble residual does not include intrinsic scatter, the age-bias could be the primary cause of the observed extra dimming of distant SNe Ia. Furthermore, we use the PAge formalism, which is a good approximation to many dark energy and modified gravity models, to do a model comparison. We find that if intrinsic scatter is included in the Hubble residual, the Lambda cold dark matter model remains a good fit. However, in a scenario without intrinsic scatter, the Lambda cold dark matter model faces a challenge.
\end{abstract}

\maketitle


\section{Introduction}\label{intro}
Supernova cosmology has been viewed as offering the clearest proof of an accelerating universe for many years. The empirical standardization of its peak luminosity is obtained by calibrating light-curve width and color~\citep{tripp1998two,phillips1999reddening,guy2007salt2}. The generality of this standardization procedure is based on an assumption that type Ia supernova (SN Ia) outbursts are triggered by a critical condition that has little to do with the galactic environment~\citep{jha2019observational}. But this assumption has been challenged. In the past decade, a series of studies have shown that the luminosity of SNe Ia may also depend on their host galaxy properties~\citep{schiavon2006deep2,hicken2009improved,sullivan2010dependence,rigault2013evidence,rigault2015confirmation,choi2014assembly,fumagalli2016ages,uddin2020carnegie,10.5303/JKAS.2019.52.5.181,briday2022accuracy}. In particular, the correlation between progenitor age and Hubble residual (HR) has been debated in recent studies~\citep{gupta2011improved,kang2016early,jha2019observational,rose2019think,rose2020evidence,kang2020early,lee2020further,rigault2020strong,zhang2021improving,lee2022evidence,wiseman2022galaxy,wiseman2023further}. This correlation is manifested by an increase in SN Ia brightness with increasing progenitor age. Since the mean age of SN Ia progenitor evolves with redshift, the systematic bias may contribute to the observed extra dimming of distant SNe Ia and should be properly subtracted for precision cosmology that aims to measure the property of dark energy.

\defcitealias{rose2019think}{R19}\defcitealias{campbell2013cosmology}{C13}\defcitealias{kang2020early}{K20}\defcitealias{huang2020supernova}{H20}\defcitealias{lee2020further}{L20}\defcitealias{lee2022evidence}{L22}\defcitealias{zhang2021improving}{Z21}

Rose et al. (ref.~\citep{rose2019think}, hereafter~\citetalias{rose2019think}) selected 102 SNe Ia at redshift of $0.05 < z < 0.2$ with a median $z = 0.14$ from the supernova sample of Campbell et al. (ref.~\citep{campbell2013cosmology}, hereafter~\citetalias{campbell2013cosmology}) to obtain ages of the local stellar population and host galaxy, respectively. The authors found a step between the HRs of younger galaxies (age $\leq 8\,\mathrm{Gyr}$) and older galaxies (age $>8\,\mathrm{Gyr}$), but there is no correlation between the age and HR of each subgroup.
Another team Kang et al. (ref.~\citep{kang2020early}, hereafter~\citetalias{kang2020early}) analyzed 51 nearby early-type host galaxies and found a correlation between age and HR with a greater slope of $\slopep= -0.051 \pm 0.022\magGyr$. If extrapolated to higher redshift, this correlation would lead to a redshift-dependent luminosity evolution. This conclusion was immediately refuted by Rose et al. (ref.~\citep{rose2020evidence}), who argued that the analysis of~\citetalias{kang2020early} was biased by several outliers (bad-quality data points). Furthermore, Huang (ref.~\citep{huang2020supernova}, hereafter~\citetalias{huang2020supernova}) pointed out that the cosmological scenario suggested by~\citetalias{kang2020early} leads to a universe that is younger than the observed oldest stars. \citetalias{huang2020supernova} proposed a Parameterization based on cosmic Age (PAge), a unified parameterization that can well approximate many dark energy and modified gravity models, to study the cosmology with supernova magnitude evolution. \citetalias{huang2020supernova} found that a decelerating universe is consistent with the supernova data when the $\slopep$ parameter is allowed to vary in the range $\lvert \slopep\rvert \lesssim 0.057\magGyr$, as suggested by~\citetalias{kang2020early}. More recently, Lee et al. (ref.~\citep{lee2020further}, hereafter~\citetalias{lee2020further}), recalculated the age-HR correlation of SN Ia data set from~\citetalias{rose2019think}  based on Gaussian mixture models, and the outcome was consistent with the result of~\citetalias{kang2020early} from early-type host galaxies.

Lee et al. (ref.~\citep{lee2022evidence}, hereafter~\citetalias{lee2022evidence}) divided the data of~\citetalias{rose2019think} into young and old groups, with a gray area in between. The width of the gray area is taken to be greater than the typical age measurement error to avoid sample contamination between two subgroups. ~\citetalias{lee2022evidence} utilized the light-curve parameters provided in~\citetalias{campbell2013cosmology} to determine $\Delta$HR. The dependence of the luminosity on the progenitor age is estimated by taking the ratio of $\Delta$HR to $\Delta\mathrm{age}$, the difference in the median local (environmental stellar population typically within a few $\mathrm{kpc}$) ages of these two groups. Combining the measured slope, which is $\slopep = -0.040\magGyr$, with mass-weighted mean age evolution of stellar population obtained from cosmic star formation history, the authors derived the redshift evolution of $\Delta$HR and claimed that there is almost no evidence of an accelerating universe from supernova data alone. 

The analysis in~\citetalias{lee2022evidence} is straightforward and intuitively easy to understand, but it may still be challenged by the same argument that the data set is contaminated by some bad-quality samples. The present work aims to advance the analysis in \citetalias{lee2022evidence}  by using a fully Bayesian approach without artificial selection or grouping of data. Our catalog includes all the 102 supernovae in the redshift range $0.05<z<0.2$ with progenitor age measurements~\citetalias{rose2019think} and the light-curve data from~\citetalias{campbell2013cosmology}. The Bayesian approach makes use of the age uncertainty information of each supernova, and therefore naturally suppresses the statistical contribution from bad-quality samples. 

This paper is organized as follows. In Section~\ref{corre} we introduce our statistical method and apply it to the supernova catalog to update the constraint on $\slopep$. In Section~\ref{COSMO}, we use the measured $\slopep$ to correct the cosmological likelihood of type Ia supernovae~\citep{scolnic2018complete} and discuss its cosmological implication. Section~\ref{discu} concludes.
\section{Correlation between Progenitor Age and Hubble Residual}\label{corre}

For the luminosity standardization of SNe Ia, the distance modulus is given by SALT2~\citep{guy2007salt2,guy2010supernova},
\begin{equation}\label{1}
\mu_{\rm SN}=m-M+\alpha x_1-\beta c,
\end{equation}
where $m$ is the apparent magnitude, i.e., $-2.5\log_{10}(x_0)$ in~\citetalias{campbell2013cosmology}, $M$ is the absolute magnitude, and $\alpha x_1$ and $\beta c$ are correction terms depending on the light-curve width ($x_1$) and color ($c$). The relative luminosity is then represented by the Hubble residual
\begin{equation}\label{2}
\rm{HR}=\mu_{\rm SN}-\mu_{\rm model},
\end{equation}
where $\mu_{\rm model}$ is the theoretical distance modulus, given by
\begin{equation}
    \mu_\mathrm {model} = 25 + 5\log_{10}\frac{d_L}{\mathrm{Mpc}}.
\end{equation} 
The luminosity distance $d_L$ as a function of cosmological redshift $z$ is specified by the cosmological model, for instance in the flat Lambda cold dark matter ($\Lambda$CDM) model
\begin{equation}
    \left.d_L(z)\right\vert_{\rm \Lambda CDM} = (1+z)\chi(z).
\end{equation} 
The comoving angular diameter distance $\chi(z)$ is given by
\begin{equation}
    \chi(z) = \frac{c}{H_0}\int_0^z \frac{\mathrm{ d}z'}{\sqrt{\Omega_m(1+z')^3 + 1 - \Omega_m}},
\end{equation}
where $H_0$ is the Hubble constant and $\Omega_m$ is the matter abundance parameter at the present epoch.

We consider a linear regression between HR and progenitor age
\begin{equation}
    \mathrm{HR} = \frac{\Delta \mathrm{HR}}{\Delta \mathrm{age}} \times \mathrm{age} + \mathrm{intercept}.
\end{equation}
The intercept term on the right-hand side can be absorbed into the definition of $M$ parameter and will be ignored hereafter. 

To properly account for the measurement error and intrinsic scatter in the regression analysis, we write the likelihood function as
\begin{equation}\label{3}
\mathcal{L} \propto \exp\left\{-\frac{1}{2}\sum_{i}\left[\frac{\left(\rm{HR}_{\emph{i}}- \frac{\Delta \mathrm{HR}}{\Delta \mathrm{age}} \times \rm{age}_{\emph{i}}\right)^2}{\sigma_i^2}+\ln{\left(2\pi\sigma_i^2\right)}\right]\right\},
\end{equation}
where ${\rm HR}_{i}=-2.5\log_{10}{x_{0i}}+\alpha x_{1i}-\beta c_i-M-\mu_{\rm model}\left(z_i\right)$ and the index $i$ runs over the 102 low-redshift SN Ia samples used here. For each SN Ia, the total uncertainty is given by 
\begin{equation}
    \begin{aligned}
        \sigma^2 &= \left(\frac{2.5\sigma_{x_0}}{x_0\ln{1}0}\right)^2+\left(\alpha\sigma_{x_1}\right)^2+\left(\beta\sigma_c\right)^2\\
        &+2\left(-\frac{2.5\alpha}{x_0 \ln{10}}\sigma_{x_0,x_1}+\frac{2.5\beta}{x_0 \ln{10}}\sigma_{x_0,c}-\alpha\beta \sigma_{x_1,c}\right)\\
        &+\left(\frac{\Delta \mathrm{HR}}{\Delta \mathrm{age}}\sigma_{\rm age}\right)^2+\sigma_{\mathrm{pec}}^2+\sigma_{\mathrm{intrinsic}}^2,
    \end{aligned}
\end{equation}
where $\sigma_{\rm intrinsic}$  is the intrinsic scatter due to other unknow factors.  The uncertainty from the peculiar velocity uncertainty $\sigma_{\rm pec} = \frac{\partial \mu_\mathrm {model} }{ \partial z}  \frac{v_\mathrm{ pec} }{ c} $ , where $v_{\rm pec}$ is set to $500{\rm km/s}$. For flat $\Lambda$CDM model, the likelihood depends on the parameters $\{\alpha,\beta,\slopep,\Omega_{m},\sigma_{\rm intrinsic},M_{\rm eff} \}$, where $M_{\rm eff} = M - 5\log_{10}\frac{H_0}{70\mathrm{km/s/Mpc}}$. 

We employ the local age of the stellar population near each SN Ia from~\citetalias{rose2019think} and corresponding light curve data ($x_0, x_1$ and $c$) from~\citetalias{campbell2013cosmology}. Then we run Monte Carlo Markov Chain (MCMC) simulations with flat priors $\alpha\in[0.001,5]$, $\beta\in[0.5,5]$, $\slopep\in[-1,1]\magGyr$, $\Omega_m\in[0,1]$, $\sigma_{\rm intrinsic} \in [0,3] $, $M_{\rm eff}\in[-40,-10]$. 

The intrinsic scatter has been included in the analyses of \citetalias{lee2020further,lee2022evidence} and Zhang et al. (ref.~\citep{zhang2021improving}, hereafter \citetalias{zhang2021improving}). However, \citetalias{lee2022evidence} noted that HR scatter of the progenitor age at a given redshift is the most likely source. The intrinsic scatter tends to weaken the strength of the detected correlations. Here we also evaluate the scenario where $\sigma_{\rm intrinsic}$=0. Table~\ref{iii} lists the 68.3\% confidence-level bounds of parameters, covering both scenarios with intrinsic scatter and without intrinsic scatter. Notably, we find at least a $3.5\sigma$ hint of nonzero $\slopep$, which indicates a correlation between the progenitor age. We further test the robustness of the constraint on $\slopep$ by randomly removing 30 supernova samples in our catalog. Repeated tests show that with the randomly selected subgroup of data, we still find $\slopep<0$ at $\gtrsim 2\sigma$ confidence level. 
\begin{table*}[h!t]
\centering
\caption{Mean value with 1$\sigma$ error. Data: 102 low-$z$ supernova samples with progenitor age information from~\citetalias{rose2019think} and corresponding light curve data from~\citetalias{campbell2013cosmology}.}
\label{iii}
\begin{tabular}{ccc}
\hline
Parameter& With intrinsic scatter& Without intrinsic scatter\\
\hline
$\alpha$&0.16 $\pm$ 0.02&0.18 $\pm$ 0.01\\
$\beta$&2.88 $\pm$ 0.18&3.20 $\pm$ 0.14\\
$\slopep$&$-0.021 \pm$ 0.006 \,$\mathrm{mag/Gyr}$&$-0.032^{+0.004}_{-0.005}$ \,$\mathrm{mag/Gyr}$\\
$M_{\rm eff}$&$-29.58\pm0.06$&$-29.51\pm0.05$\\
$\sigma_{\rm intrinsic}$ &0.108 $\pm$ 0.015&-\\
$\Omega_m$ & $0.46^{+0.19}_{-0.25}$& $0.56\pm 0.19$\\
\hline
\end{tabular}
\end{table*}
\begin{figure}[b]
\centering
\includegraphics[width=0.46\textwidth]{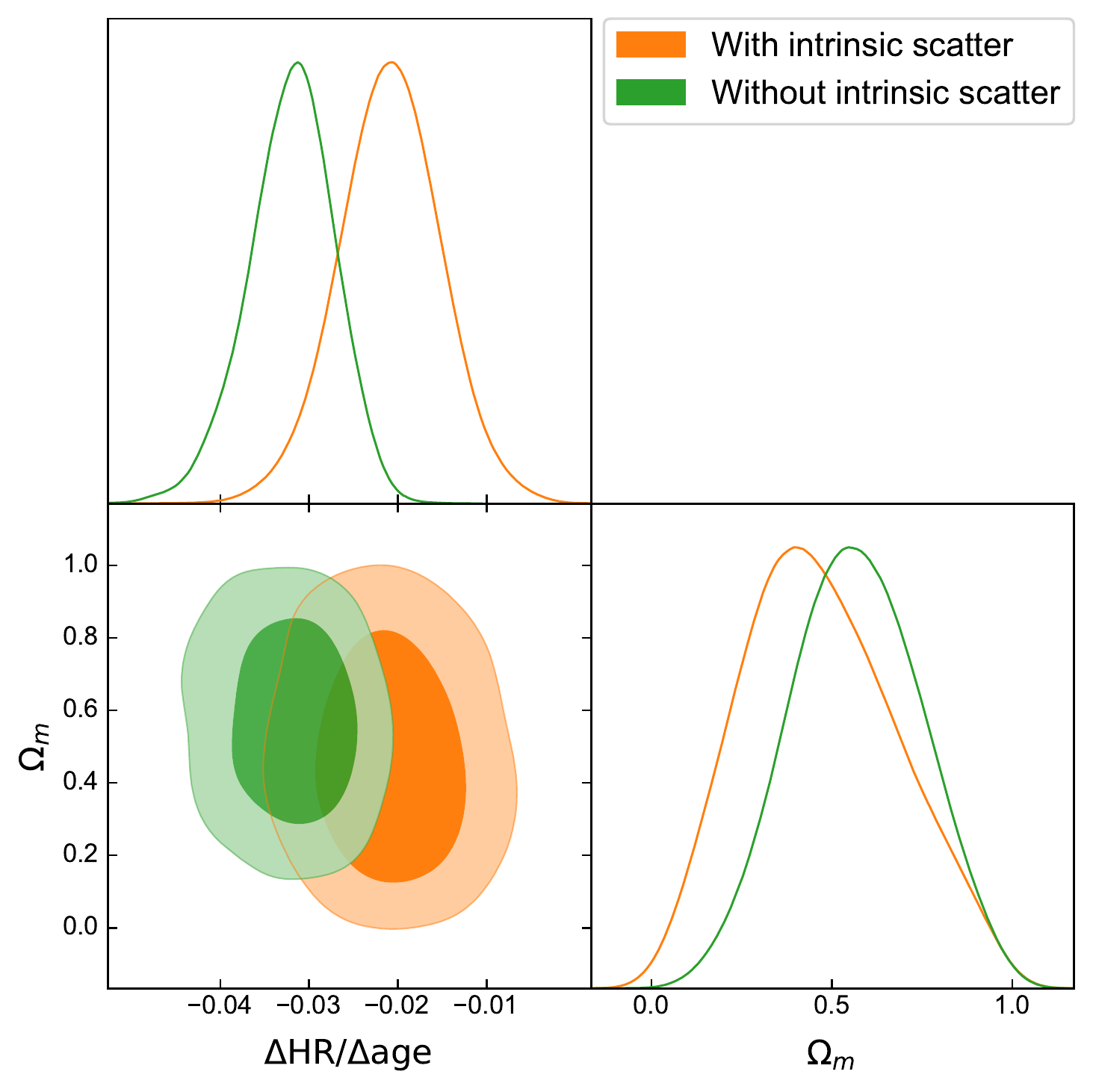}	  
\caption{The marginalized distributions of $\slopep$ and $\Omega_m$. Inner and outer contours correspond to 68.3\% and 95.4\% confidence levels, respectively. The orange and green regions represent situations involving intrinsic scatter and no intrinsic scatter, respectively.}
\label{slope-Om}
\end{figure}
\begin{figure}[b]
\centering
\includegraphics[width=0.46\textwidth]{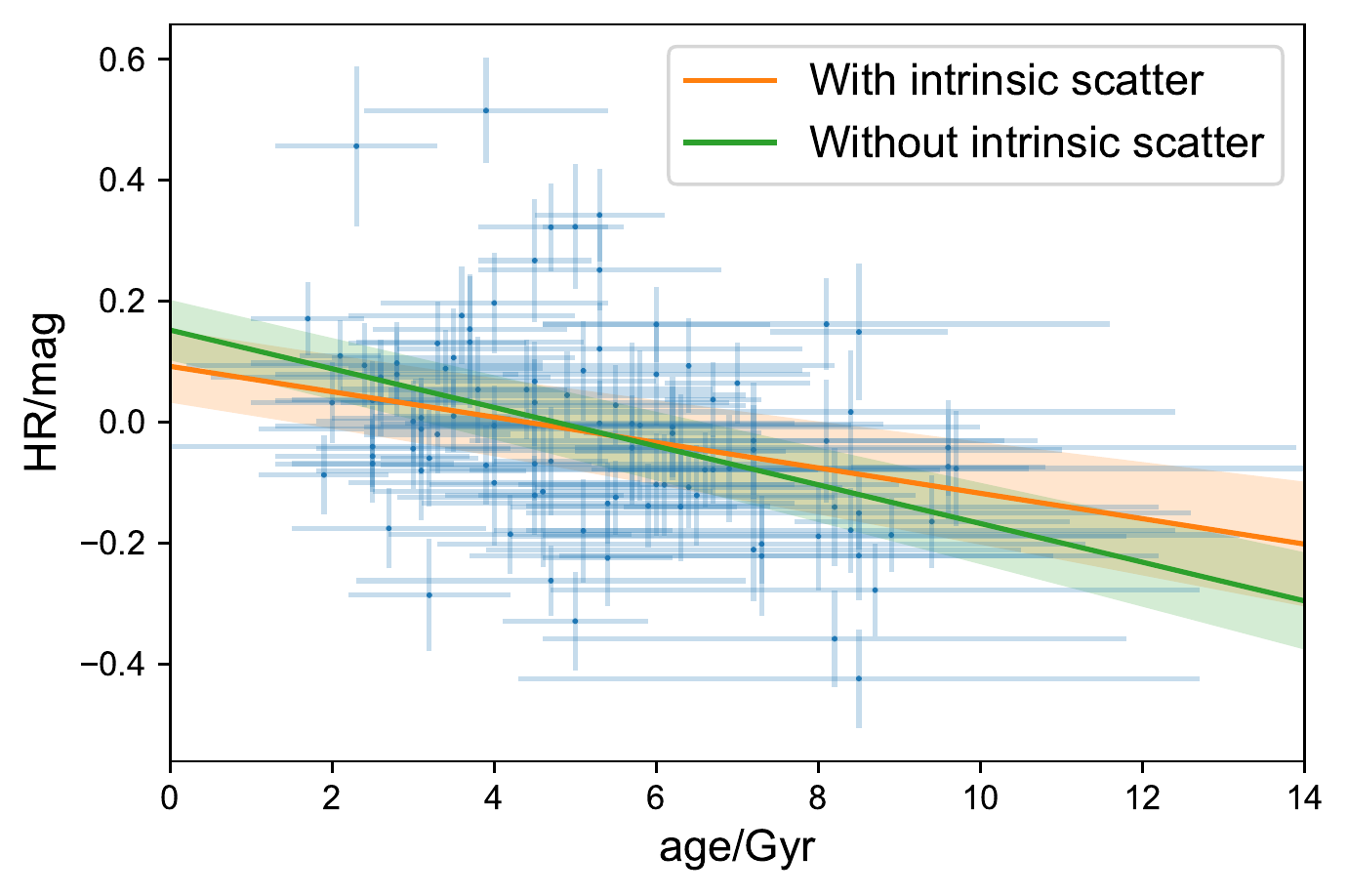}  
\caption{The age–luminosity relation. The data points are 102 SNe Ia in \citetalias{rose2019think}. The orange and green lines, corresponding to cases with intrinsic scatter and without intrinsic scatter respectively, represent regression fittings with the mean value of $\slopep$ given in Table~\ref{iiii}. The shaded area represents the 1$\sigma$ level region.}
\label{slope}
\end{figure}
As shown in Figure~\ref{slope-Om}, there is no noticeable degeneracy between $\slopep$ and $\Omega_m$. The preference of a nonzero $\slopep$ is then, in the $\Lambda$CDM framework, cosmology-independent. This is somewhat expected because the calibration of $\slopep$ only involves low-redshift ($z\le 0.2$ here) supernova data, which is insensitive to cosmology. For the same reason, the analysis here has almost no constraining power on cosmological parameters. To validate this claim, we directly constrained the parameters using the \citetalias{rose2019think} HR dataset. The HR values in the data set are ascertained through \citetalias{campbell2013cosmology}, with the stretch parameter $\alpha$ being adjusted to $0.16$. The constrained results are shown in Table~\ref{iiii}. The Hubble residual is directly visualized in Figure~\ref{slope}. We find that $\slopep$ almost remains unaffected.
\begin{table*}[h!t]
\centering
\caption{Mean value with 1$\sigma$ error. Data: 102 low-$z$ supernova samples with progenitor age information and corresponding HR value from~\citetalias{rose2019think}.}
\label{iiii}
\begin{tabular}{ccc}
\hline
Parameter& With intrinsic scatter& Without intrinsic scatter\\
\hline
$\slopep$&$-0.021 \pm$ 0.006 \,$\mathrm{mag/Gyr}$&$-0.033^{+0.004}_{-0.005}$ \,$\mathrm{mag/Gyr}$\\
$\mathrm{Intercept}$&$0.092\pm0.034$&$0.152^{+0.02}_{-0.03}$\\
$\sigma_{\rm intrinsic}$ &$0.117^{+0.013}_{-0.015}$&- \\
\hline
\end{tabular}
\end{table*}
In the next section we study cosmology with the Pantheon supernova catalog that covers a much broader redshift range $0<z<1.5$~\citep{scolnic2018complete}, based on the assumption that the HR-age relation also applies to high-redshift SNe Ia. 
\section{Cosmology Analysis}\label{COSMO}

The age information of individual samples in the Pantheon supernova catalog is not available. We instead use the mean age of the stellar population as a function of redshift, derived from the theory of cosmic star formation history and tabulated in Table 1 of~\citetalias{lee2022evidence}. 
We utilize the heliocentric redshift $z_{\rm hel}$, the CMB restframe redshift $z_{\rm cmb}$ and the covariance matrix of distance modulus provided by Pantheon supernova catalog. The luminosity distance of SN Ia can be expressed as $d_L=(1+z_\mathrm{hel})\chi(z_\mathrm{ cmb})$.  The nuisance parameter $M$ is marginalized using the method of ref.~\citep{conley2010supernova}. Additionally, we removed six supernovae with redshifts $z>1.5$ that are out of the tabulated redshift range and revised the standard supernova likelihood by adding a $\slopep \times \mathrm{age}$ term to the distance modulus, where $\mathrm{age}$ is the mean age of the stellar population, and apply a flat prior on $\slopep\in [-0.1, 0.1]\,\mathrm{mag/Gyr}$.
\begin{figure}[b]
\centering
\includegraphics[width=0.46\textwidth]{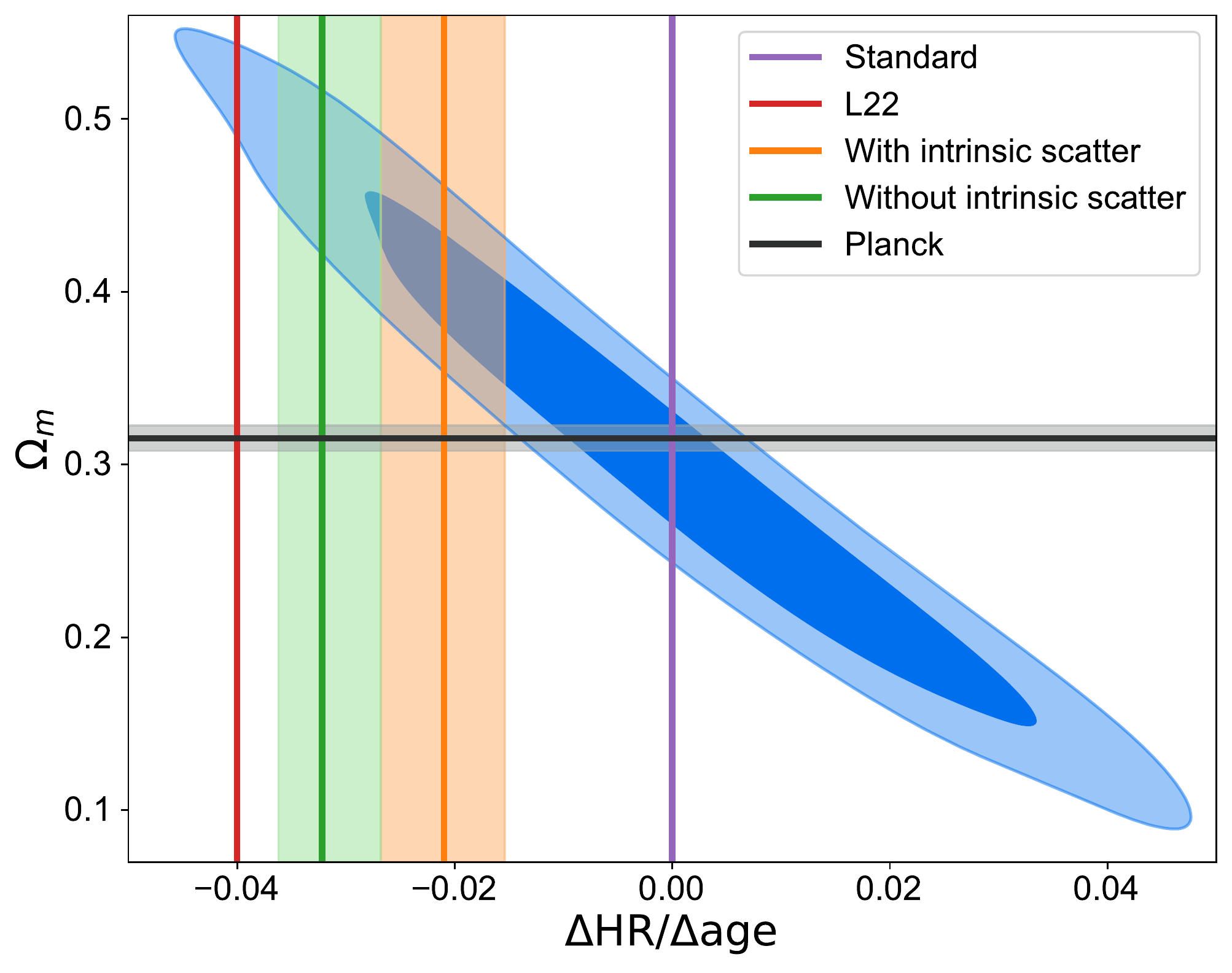}
\caption{ The marginalized $68.3\%$ (inner blue contour) and $95.4\%$ (outer blue contour) confidence level constraints with Pantheon SN Ia data set and a flat prior on $\slopep$. The horizontal black band stands for Planck constraint on matter abundance ($\Omega_m=0.315 \pm 0.007$). The vertical bands are four cases of supernova magnitude evolution: (i) the standard assumption of no HR-age dependence ($\slopep = 0$, purple); (ii) the \citetalias{lee2022evidence} result $\slopep=-0.04\magGyr$ (red); (iii) $\slopep = -0.021\pm 0.006\magGyr$ (orange), which was identified in this research including intrinsic scatter; (iv) $\slopep = -0.032^{+0.004}_{-0.005}$ (green), as found in this study without intrinsic scatter. We label these four cases on $\slopep$ as ``Standard'', ``\citetalias{lee2022evidence}'', ``With intrinsic scatter'' and ``Without intrinsic scatter'', respectively. }
\label{lcdm}
\end{figure} 

\begin{figure*}[htbp]
\centering
\includegraphics[width=0.46\textwidth]{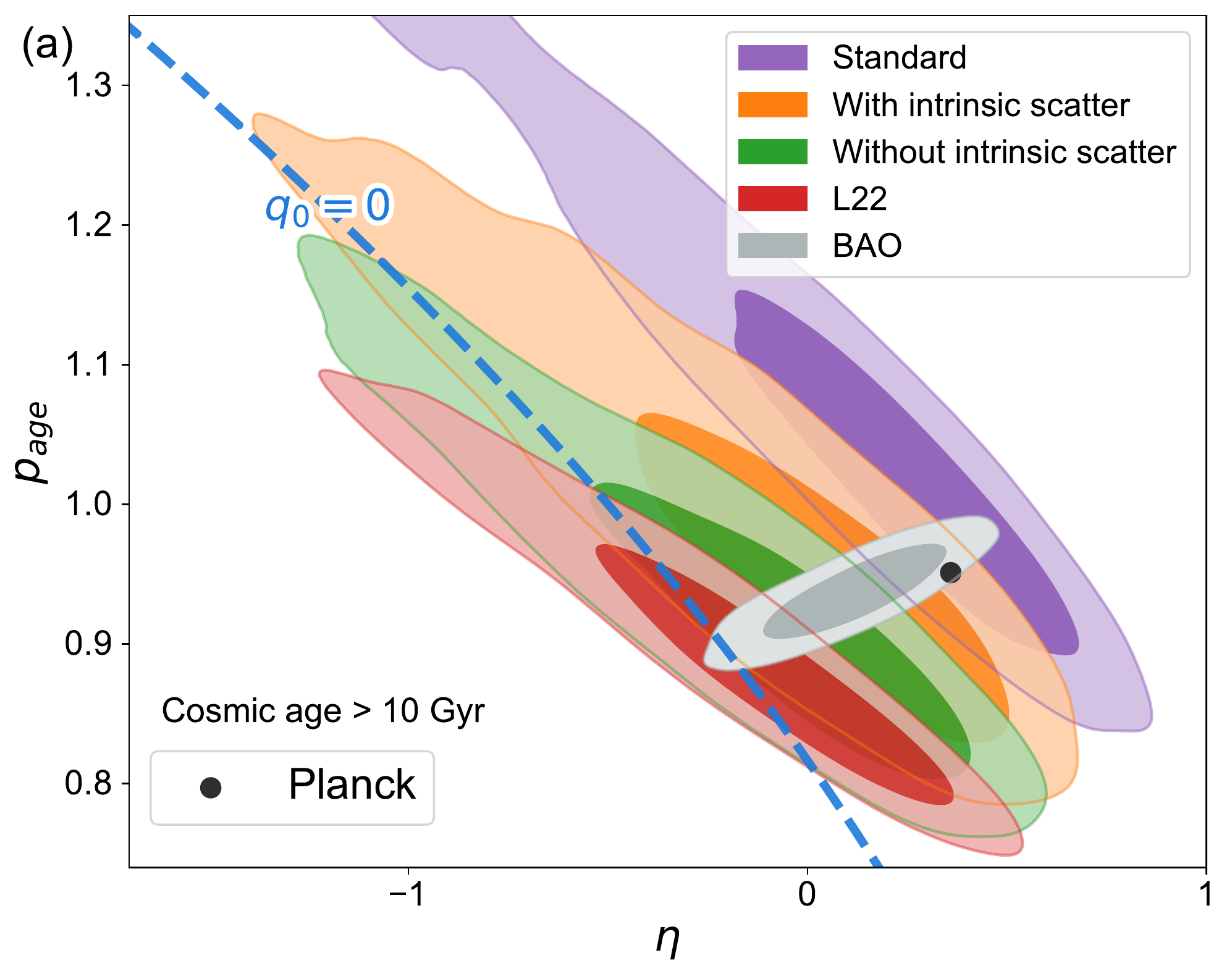}
\includegraphics[width=0.46\textwidth]{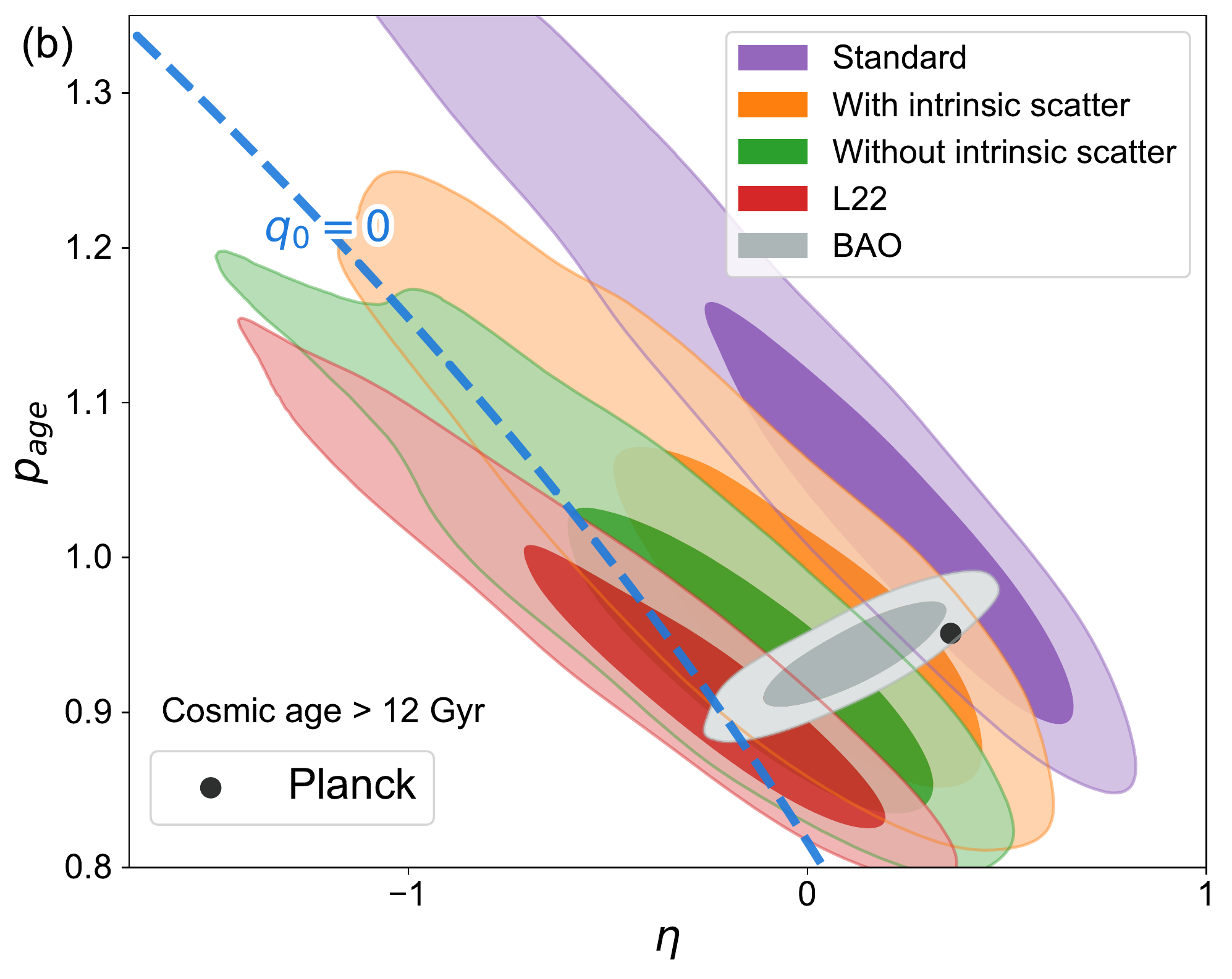}
\caption{The $1\sigma$ and $2\sigma$ confidence level contours of the flat PAge approximation for the supernova data with four priors on $\slopep$ and for the BAO data. Figure (a) and Figure (b) respectively depict the scenarios with a lower age limit for the universe of 10 Gyr and 12 Gyr. The black dot $(\eta, p_{\rm age}) = (0.359, 0.951)$ is a good approximation to the Planck best-fit $\Lambda$CDM model ($\Omega_{m} \approx 0.315$).  The dashed blue $q_0=0$ line is the critical line between cosmic acceleration (upper right region) and deceleration (lower left region).}
\label{page}
\end{figure*} 
\begin{figure*}[htbp]
\centering
\includegraphics[width=0.23\textwidth]{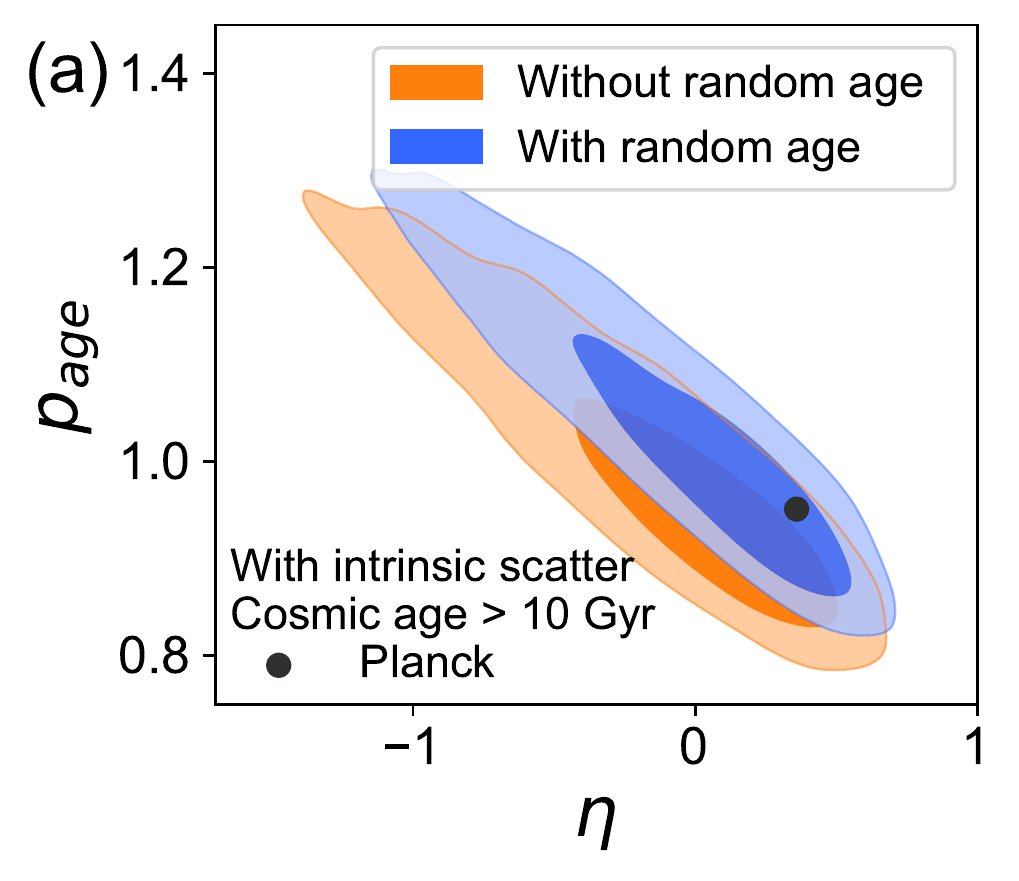}
\includegraphics[width=0.23\textwidth]{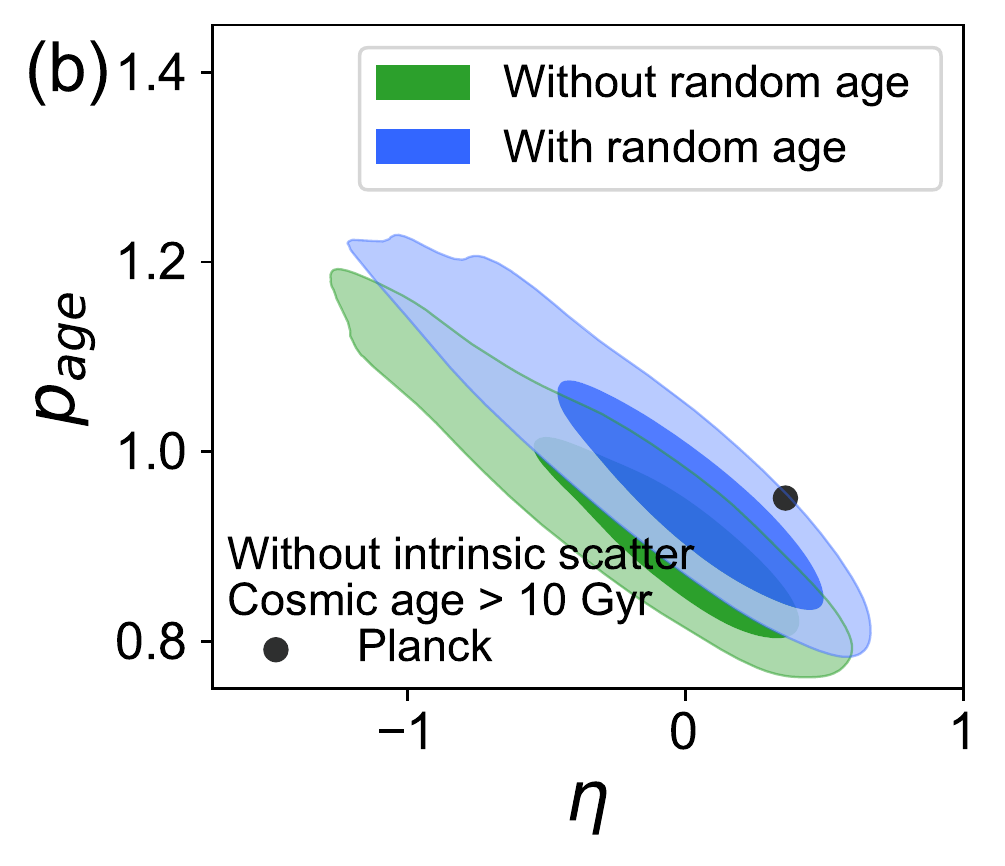}
\includegraphics[width=0.23\textwidth]{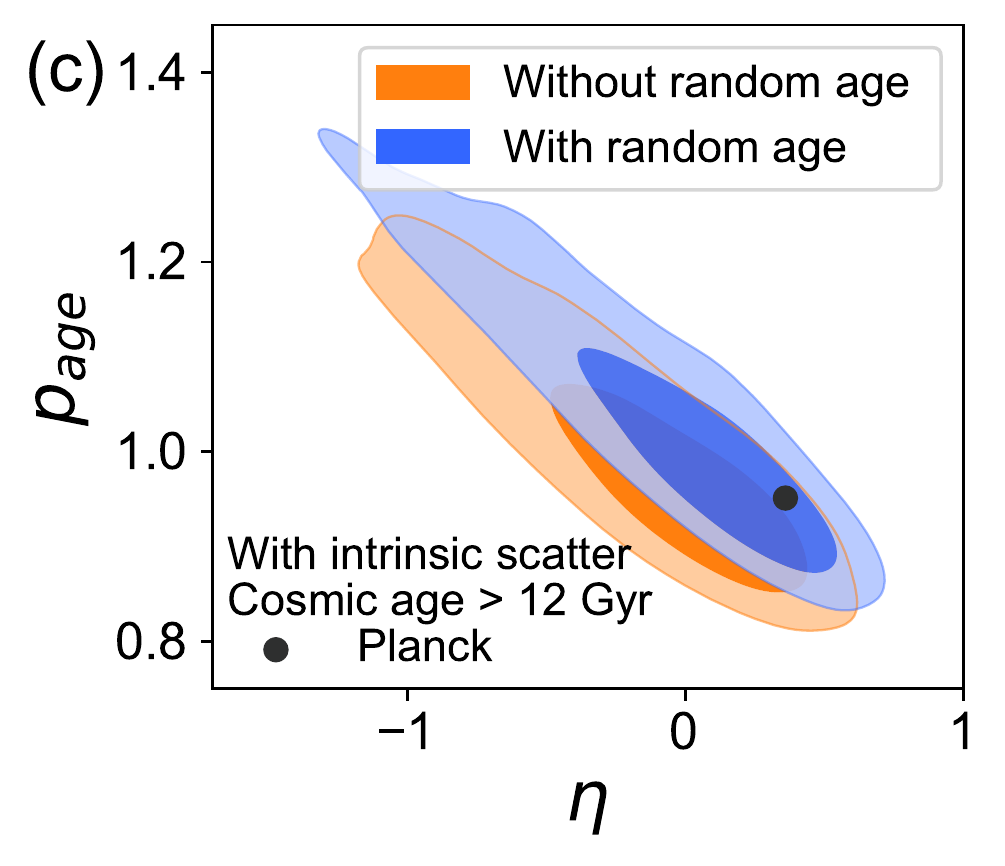}
\includegraphics[width=0.23\textwidth]{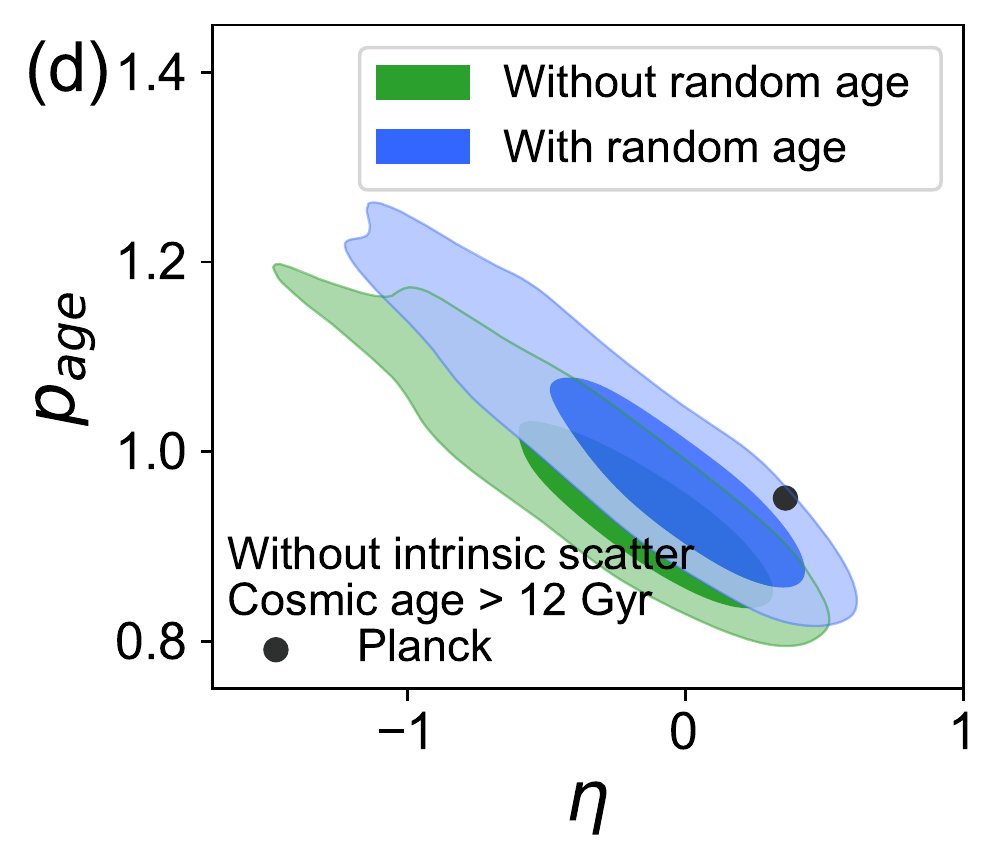}
\caption{The $1\sigma$ and $2\sigma$ confidence level contours of the flat PAge approximation for the supernova data shown in Figure (a) and (c) with the prior $\slopep = -0.021\pm 0.006\magGyr$, and in Figure (b) and (d) with the prior $\slopep = -0.032\pm 0.005\magGyr$. Figure (a) and (b) depict scenarios with a lower age limit for the universe of 10 Gyr, while Figure (c) and (d) represent scenarios with a lower age limit of 12 Gyr. The inner blue contours are the cases where $-1 \sim 1$ Gyr age fluctuation is randomly added to each supernova. The black dot $(\eta, p_{\rm age}) = (0.359, 0.951)$ is a good approximation to the Planck best-fit $\Lambda$CDM model ($\Omega_{m} \approx 0.315$).}
\label{page_random}
\end{figure*} 

The marginalized constraints on $\Omega_m$ and $\slopep$ are compared with four cases of supernova magnitude evolution: (i) the standard assumption of no HR-age dependence ($\slopep = 0$); (ii) $\slopep=-0.04\magGyr$ that was found in~\citetalias{lee2022evidence}; (iii) $\slopep = -0.021\pm 0.006\magGyr$ (Gaussian error) that was found in this work with intrinsic scatter, and (iv) $\slopep = -0.032\pm 0.005\magGyr$ (Gaussian error) for the case without intrinsic scatter. The result is visually shown in Figure~\ref{lcdm}, which is a contour with Pantheon SN Ia data set and a flat prior on $\slopep$. Compared to the standard ($\slopep=0$) case, the HR-age dependence in the case with intrinsic scatter ($\slopep = -0.021\pm 0.006\magGyr$) leads to a larger $\Omega_m = 0.392 \pm 0.037$ that is in mild ($\sim 2\sigma$) tension with the concordance model ($\Omega_m = 0.315\pm 0.007$) ~\citep{Planck18}. For the case without intrinsic scatter ($\slopep = -0.032\pm 0.005\magGyr$), the matter abundance further increases to $\Omega_m = 0.480^{+0.040}_{-0.047}$, which is in $\sim 3.5\sigma$  tension with the concordance model\footnote{The matter abundance $\Omega_m$ here is constrained by the Pantheon SN Ia data set, while $\Omega_m$ in Table~\ref{iii} is constrained by the 102 low-redshift SN Ia samples, so they are different.}. If we accept the assumption that the HR-age dependence can be extended to high redshift, the tension with the Planck standard cosmology could be a hint of new physics beyond the flat $\Lambda$CDM model. 

To test whether a non-standard cosmology can fit the data much better, we use the PAge approximation, which provides a simple and almost model-independent approximation to $\Lambda$CDM and many beyond-$\Lambda$CDM models~\citep{huang2020supernova, luo2020reaffirming, MAPAge, GRB2021PAge,huang2022s, Huang2022PAge, cai2022no2, cai2022no1, li2022redshift}. Model comparison can be easily done in the PAge parameter space with one single MCMC run. The PAge approximation is formulated as 
\begin{equation}\label{4}
\frac{H}{H_0}=1+\frac{2}{3}\left (1-\eta\frac{H_0t}{p_{\mathrm{age}}}\right )\left (\frac{1}{H_0t}-\frac{1}{p_{\mathrm{age}}}\right ),
\end{equation}
where $t$ is the cosmological time. $p_{\rm age}=H_0t_0$ is the product of the Hubble constant $H_0$ and the current age of the universe $t_0$. The phenomenological parameter $\eta$ can be regarded as a fitting parameter that represents the deviation from an Einstein-de Sitter Universe. Studying the representation of a specific model in the PAge parameter space requires (i) the product of the current Hubble constant ($H_0$) and cosmic age ($t_0$), and (ii) the present-day deceleration parameter ($q_0$). Subsequently, we can obtain the PAge parameter $p_{age} = H_0 t_0$ and $\eta = 1 - \frac{3}{2} p_{age}^2(1 + q_0)$. The Planck concordance cosmology ($\Omega_m\approx 0.315$) is well approximated by the PAge model with $p_{\rm age} = 0.951$ and $\eta= 0.359$.

Because PAge is a phenomenological parameterization for the late universe, we are not supposed to apply PAge to cosmic microwave background data, whose anisotropy involves the evolution of cosmological perturbations from the early universe. For comparison with external data, we instead use the baryon acoustic oscillations (BAOs). The BAO data set includes the latest measurements from the Sloan Digital Sky Survey (SDSS)-IV\footnote{https://svn.sdss.org/public/data/eboss/DR16cosmo/\\
tags/v1\_0\_0/likelihoods
/BAO-only/}~\citep{alam2021completed}, the Six-degree Field Galaxy Survey (6dFGS)~\citep{beutler20116df}, and the Dark Energy Survey Year 3 (DES Y3) final release~\citep{abbott2022dark}. 

The mean ages in Table 1 of ~\citetalias{lee2022evidence} are computed with a reference $\Lambda$CDM model. Applying these ages to models that are beyond, but in the proximity of $\Lambda$CDM is an approximate method. To avoid running into parameter space that is too far away from the concordance model and hence the approximation may fail, for all cases we apply a lower limit for the age of the universe and a flat prior for the Hubble constant $H_0 \in [65,75]\,\mathrm{km/s/Mpc}$ to restrict the parameter space. For the lower limit age of the universe, we consider the 12 Gyr age provided by metal-poor globular clusters (GC)~\citep{vandenberg2014three,catelan2017ages,sahlholdt2019benchmark}. However, due to the uncertainty in the absolute age determination of GCs, we also investigate the impact of a lower age (10 Gyr) of metal-poor GCs. 

We then run MCMC calculations by replacing the flat prior on $\slopep$ with the four cases of supernova magnitude evolution, the ``standard'' case with fixed $\slopep=0$, the ``\citetalias{lee2022evidence}'' case with fixed $\slopep=-0.04\magGyr$, ``with intrinsic scatter'' case $\slopep=-0.021\pm 0.006\magGyr$, and ``without intrinsic scatter'' case $\slopep=-0.032\pm 0.005\magGyr$. The marginalized constraints on PAge parameters $p_{\rm age}$ and $\eta$ are shown in Figure~\ref{page}. Compared to the ``\citetalias{lee2022evidence}'' case where a decelerating universe is more favored, the updated constraint on $\slopep$ including intrinsic scatter in this work reconciles cosmic acceleration at $\sim 2\sigma$ level. The Planck best-fit $\Lambda$CDM model, well approximated by the $(\eta, p_{\rm age}) = (0.359, 0.951)$ point, is in $1.5\sigma$ tension with the best-fit parameters when considering the intrinsic scatter. While for the case without intrinsic scatter, we find $\sim 1\sigma$ hint of cosmic acceleration, and $\sim 3 \sigma$ tension to the Planck best-fit $\Lambda$CDM model. For a given redshift, the age uncertainties for different SNe Ia may have implications. Figure~\ref{page_random} shows that after adding a random age fluctuation of $-1 \sim 1$Gyr to each supernova, the tension between this work and CMB is further reduced. 

In both cases of the Hubble residual including and excluding intrinsic scatter, we conduct a comparison of the difference in the best-fit $\chi^2$ values, which are defined as $-2\ln \mathcal{L}$, between the MCMC results for the $\Lambda$CDM model and the PAge model. When we examine the case with intrinsic scatter, the difference in the best-fit $\chi^2$ is less than 0.3, suggesting no significant evidence supporting a model beyond $\Lambda$CDM model. However, for the case without intrinsic scatter, the best-fit $\chi^2$ for the Page model is 6.2 lower than that for the $\Lambda$CDM model, suggesting a challenge to the $\Lambda$CDM model.

\section{Discussion and Conclusion}\label{discu}
The usage of standardized peak luminosity of type Ia supernova as a standard candle has been recently challenged by a series of works that claim the detection of correlations between the standardized luminosity and properties of the host galaxies. In particular,  for low-redshift supernovae whose environmental stellar population can be resolved,~\citetalias{lee2022evidence} found a dependence of the Hubble residual on the age of its environmental stellar population. The analysis in~\citetalias{lee2022evidence} does not make use of the age uncertainties, and therefore may be contaminated by bad-quality samples. The present work advances the analysis in~\citetalias{lee2022evidence} by including the age uncertainties in a fully Bayesian inference framework and finds a significantly weaker dependence of the Hubble residual on the progenitor age, $\slopep=-0.021\pm 0.006\magGyr$ with intrinsic scatter around 0.11.

\begin{figure}[htbp]
\centering
\includegraphics[width=0.46\textwidth]{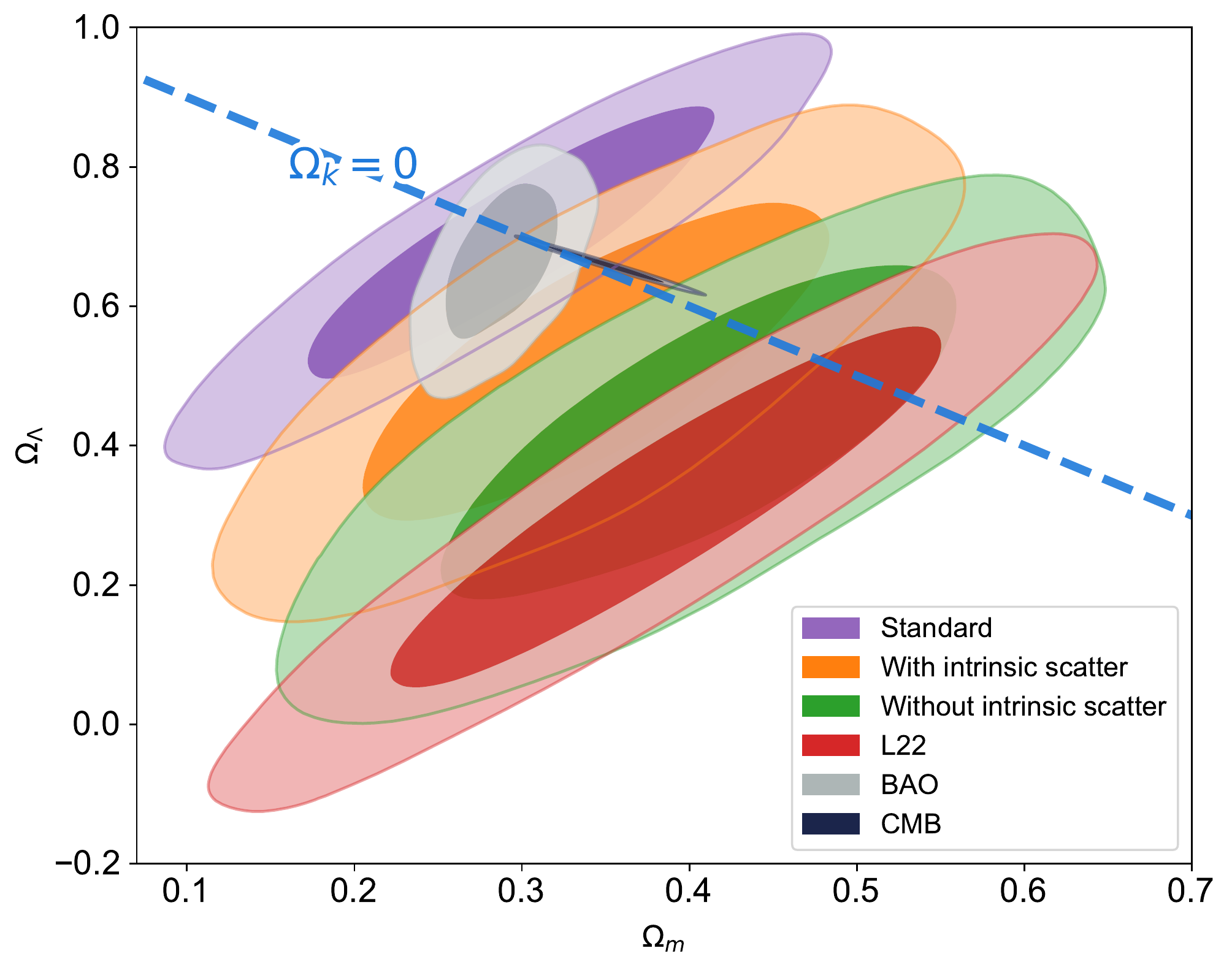}
\caption{ The marginalized $68.3\%$ and $95.4\%$ 
confidence level constraints for the $\Omega_m$ and $\Omega_{\Lambda}$ cosmological parameters under the non-flat $\Lambda$CDM model. For this diagram, SN Ia and BAO data sets are the same as Figure~\ref{page}. The CMB data are from the base\_omegak\_plikHM\_TTTEEE\_low\_lowE\_lensing chain provided by Planck~\citep{Planck18}.}
\label{ocdm}
\end{figure}

With a bold assumption that the progenitor age dependence can be extended to high redshift supernovae, we studied the impact of the supernova luminosity evolution on cosmology. We find that if Hubble residual includes the intrinsic scatter, $\Lambda$CDM model remains a good fit for the supernova data, but a high $\Omega_m$ value is preferred ($\sim 2\sigma$ larger than the Planck results). In contrast, the~\citetalias{lee2022evidence} result ($\slopep=-0.04\magGyr$) almost excludes (in $\sim 4\sigma$ tension with) the Planck concordance model. In addition, under the non-flat $\Lambda$CDM model, both CMB and BAO are consistent with the result in this work if we consider the intrinsic scatter (Figure~\ref{ocdm}).

\citetalias{lee2022evidence} is based upon previous research, such as \citetalias{kang2020early}, \citetalias{lee2020further}, and \citetalias{zhang2021improving}, which all thoroughly considered the uncertainty of progenitor ages. \citetalias{kang2020early} claimed to have discovered a correlation between the ages of 34 early-type host galaxies and the peak luminosity of Type Ia supernovae, with a slope $\slopep = -0.051 \pm 0.022 \magGyr$. However, \citetalias{rose2019think} pointed out that due to poor sampling of the light curves of Type Ia supernovae, \citetalias{kang2020early} overestimated the slope. \citetalias{lee2020further} and \citetalias{zhang2021improving} conducted their studies using the same data source as in this paper, namely the \citetalias{rose2019think} HR dataset. \citetalias{lee2020further} presented a relatively steep value of $\slopep =  -0.057 \pm 0.016 \magGyr$. We have found that the HR values used by \citetalias{lee2020further} were based on an $\alpha$ value of $0.22$. However, \citetalias{campbell2013cosmology} and \citetalias{rose2019think} indicated that this $\alpha$ value was larger than the typical value. Consequently, this larger $\alpha$ amplified the results of \citetalias{lee2020further}. By changing the $\alpha$ value to $0.16$ and replicating the work of \citetalias{lee2020further} using the LINMIX package, we obtained $\slopep =  -0.032 \pm 0.017 \magGyr$, which differs from our results by only a $0.7\sigma$ level. This value is close to the one obtained by \citetalias{zhang2021improving}, $\slopep =  -0.035 \pm 0.007 \magGyr$, whose posterior sampling method adopts the full posterior for the age error instead of the Gaussian error. The tension between the results of \citetalias{zhang2021improving} and our findings is at a $1.5\sigma$ level.

Another possible interpretation of the luminosity evolution is that the SNe Ia survey strategy favors large-mass galaxies, and the median age of the host galaxies of Pantheon catalog may be biased by the selection effect.  Moreover, Pantheon catalog and the low-redshift catalog with age information are two different catalogs and may follow different statistics. A more detailed calculation of the age distribution of the host galaxies of the Pantheon catalog is beyond the scope of the present work. We leave exploration along this direction as our future work. While the impact of the selection effect remains uncertain, we argue that the luminosity evolution problem remains a possible systematic error and is worthy of in-depth study for future high-precision supernova cosmology and more detailed comparison with other cosmological probes in the near future~\cite{ade2019simons,liu2022forecasts,miao2023cosmological}. 

\section*{Acknowledgements}
This work was supported by the National SKA Program of China (Grant No. 2020SKA0110402), the National Natural Science Foundation of China (NSFC) (Grant No. 12073088), the National Key R\&D Program of China (Grant No. 2020YFC2201600), and the Guangdong Major Project of Basic and Applied Basic Research (Grant No. 2019B030302001).

\bibliography{biblio} 

\begin{thebibliography}{50}%
\makeatletter
\providecommand \@ifxundefined [1]{%
 \@ifx{#1\undefined}
}%
\providecommand \@ifnum [1]{%
 \ifnum #1\expandafter \@firstoftwo
 \else \expandafter \@secondoftwo
 \fi
}%
\providecommand \@ifx [1]{%
 \ifx #1\expandafter \@firstoftwo
 \else \expandafter \@secondoftwo
 \fi
}%
\providecommand \natexlab [1]{#1}%
\providecommand \enquote  [1]{``#1''}%
\providecommand \bibnamefont  [1]{#1}%
\providecommand \bibfnamefont [1]{#1}%
\providecommand \citenamefont [1]{#1}%
\providecommand \href@noop [0]{\@secondoftwo}%
\providecommand \href [0]{\begingroup \@sanitize@url \@href}%
\providecommand \@href[1]{\@@startlink{#1}\@@href}%
\providecommand \@@href[1]{\endgroup#1\@@endlink}%
\providecommand \@sanitize@url [0]{\catcode `\\12\catcode `\$12\catcode `\&12\catcode `\#12\catcode `\^12\catcode `\_12\catcode `\%12\relax}%
\providecommand \@@startlink[1]{}%
\providecommand \@@endlink[0]{}%
\providecommand \url  [0]{\begingroup\@sanitize@url \@url }%
\providecommand \@url [1]{\endgroup\@href {#1}{\urlprefix }}%
\providecommand \urlprefix  [0]{URL }%
\providecommand \Eprint [0]{\href }%
\providecommand \doibase [0]{https://doi.org/}%
\providecommand \selectlanguage [0]{\@gobble}%
\providecommand \bibinfo  [0]{\@secondoftwo}%
\providecommand \bibfield  [0]{\@secondoftwo}%
\providecommand \translation [1]{[#1]}%
\providecommand \BibitemOpen [0]{}%
\providecommand \bibitemStop [0]{}%
\providecommand \bibitemNoStop [0]{.\EOS\space}%
\providecommand \EOS [0]{\spacefactor3000\relax}%
\providecommand \BibitemShut  [1]{\csname bibitem#1\endcsname}%
\let\auto@bib@innerbib\@empty
\bibitem [{\citenamefont {Tripp}(1998)}]{tripp1998two}%
  \BibitemOpen
  \bibfield  {author} {\bibinfo {author} {\bibfnamefont {R.}~\bibnamefont {Tripp}},\ }\bibfield  {title} {\bibinfo {title} {A two-parameter luminosity correction for type ia supernovae},\ }\href@noop {} {\bibfield  {journal} {\bibinfo  {journal} {Astronomy and Astrophysics}\ }\textbf {\bibinfo {volume} {331}},\ \bibinfo {pages} {815} (\bibinfo {year} {1998})}\BibitemShut {NoStop}%
\bibitem [{\citenamefont {Phillips}\ \emph {et~al.}(1999)\citenamefont {Phillips}, \citenamefont {Lira}, \citenamefont {Suntzeff}, \citenamefont {Schommer}, \citenamefont {Hamuy},\ and\ \citenamefont {Maza}}]{phillips1999reddening}%
  \BibitemOpen
  \bibfield  {author} {\bibinfo {author} {\bibfnamefont {M.}~\bibnamefont {Phillips}}, \bibinfo {author} {\bibfnamefont {P.}~\bibnamefont {Lira}}, \bibinfo {author} {\bibfnamefont {N.~B.}\ \bibnamefont {Suntzeff}}, \bibinfo {author} {\bibfnamefont {R.}~\bibnamefont {Schommer}}, \bibinfo {author} {\bibfnamefont {M.}~\bibnamefont {Hamuy}},\ and\ \bibinfo {author} {\bibfnamefont {J.}~\bibnamefont {Maza}},\ }\bibfield  {title} {\bibinfo {title} {The reddening-free decline rate versus luminosity relationship for type ia supernovae},\ }\href@noop {} {\bibfield  {journal} {\bibinfo  {journal} {The Astronomical Journal}\ }\textbf {\bibinfo {volume} {118}},\ \bibinfo {pages} {1766} (\bibinfo {year} {1999})}\BibitemShut {NoStop}%
\bibitem [{\citenamefont {Guy}\ \emph {et~al.}(2007)\citenamefont {Guy}, \citenamefont {Astier}, \citenamefont {Baumont}, \citenamefont {Hardin}, \citenamefont {Pain}, \citenamefont {Regnault}, \citenamefont {Basa}, \citenamefont {Carlberg}, \citenamefont {Conley}, \citenamefont {Fabbro} \emph {et~al.}}]{guy2007salt2}%
  \BibitemOpen
  \bibfield  {author} {\bibinfo {author} {\bibfnamefont {J.}~\bibnamefont {Guy}}, \bibinfo {author} {\bibfnamefont {P.}~\bibnamefont {Astier}}, \bibinfo {author} {\bibfnamefont {S.}~\bibnamefont {Baumont}}, \bibinfo {author} {\bibfnamefont {D.}~\bibnamefont {Hardin}}, \bibinfo {author} {\bibfnamefont {R.}~\bibnamefont {Pain}}, \bibinfo {author} {\bibfnamefont {N.}~\bibnamefont {Regnault}}, \bibinfo {author} {\bibfnamefont {S.}~\bibnamefont {Basa}}, \bibinfo {author} {\bibfnamefont {R.}~\bibnamefont {Carlberg}}, \bibinfo {author} {\bibfnamefont {A.}~\bibnamefont {Conley}}, \bibinfo {author} {\bibfnamefont {S.}~\bibnamefont {Fabbro}}, \emph {et~al.},\ }\bibfield  {title} {\bibinfo {title} {Salt2: using distant supernovae to improve the use of type ia supernovae as distance indicators},\ }\href@noop {} {\bibfield  {journal} {\bibinfo  {journal} {Astronomy \& Astrophysics}\ }\textbf {\bibinfo {volume} {466}},\ \bibinfo {pages} {11} (\bibinfo {year} {2007})}\BibitemShut {NoStop}%
\bibitem [{\citenamefont {Jha}\ \emph {et~al.}(2019)\citenamefont {Jha}, \citenamefont {Maguire},\ and\ \citenamefont {Sullivan}}]{jha2019observational}%
  \BibitemOpen
  \bibfield  {author} {\bibinfo {author} {\bibfnamefont {S.~W.}\ \bibnamefont {Jha}}, \bibinfo {author} {\bibfnamefont {K.}~\bibnamefont {Maguire}},\ and\ \bibinfo {author} {\bibfnamefont {M.}~\bibnamefont {Sullivan}},\ }\bibfield  {title} {\bibinfo {title} {Observational properties of thermonuclear supernovae},\ }\href@noop {} {\bibfield  {journal} {\bibinfo  {journal} {Nature Astronomy}\ }\textbf {\bibinfo {volume} {3}},\ \bibinfo {pages} {706} (\bibinfo {year} {2019})}\BibitemShut {NoStop}%
\bibitem [{\citenamefont {Schiavon}\ \emph {et~al.}(2006)\citenamefont {Schiavon}, \citenamefont {Faber}, \citenamefont {Konidaris}, \citenamefont {Graves}, \citenamefont {Willmer}, \citenamefont {Weiner}, \citenamefont {Coil}, \citenamefont {Cooper}, \citenamefont {Davis}, \citenamefont {Harker} \emph {et~al.}}]{schiavon2006deep2}%
  \BibitemOpen
  \bibfield  {author} {\bibinfo {author} {\bibfnamefont {R.~P.}\ \bibnamefont {Schiavon}}, \bibinfo {author} {\bibfnamefont {S.}~\bibnamefont {Faber}}, \bibinfo {author} {\bibfnamefont {N.}~\bibnamefont {Konidaris}}, \bibinfo {author} {\bibfnamefont {G.}~\bibnamefont {Graves}}, \bibinfo {author} {\bibfnamefont {C.~N.}\ \bibnamefont {Willmer}}, \bibinfo {author} {\bibfnamefont {B.~J.}\ \bibnamefont {Weiner}}, \bibinfo {author} {\bibfnamefont {A.~L.}\ \bibnamefont {Coil}}, \bibinfo {author} {\bibfnamefont {M.~C.}\ \bibnamefont {Cooper}}, \bibinfo {author} {\bibfnamefont {M.}~\bibnamefont {Davis}}, \bibinfo {author} {\bibfnamefont {J.}~\bibnamefont {Harker}}, \emph {et~al.},\ }\bibfield  {title} {\bibinfo {title} {The deep2 galaxy redshift survey: Mean ages and metallicities of red field galaxies at z\~{} 0.9 from stacked keck deimos spectra},\ }\href@noop {} {\bibfield  {journal} {\bibinfo  {journal} {The Astrophysical Journal}\ }\textbf {\bibinfo {volume} {651}},\ \bibinfo {pages} {L93} (\bibinfo {year}
  {2006})}\BibitemShut {NoStop}%
\bibitem [{\citenamefont {Hicken}\ \emph {et~al.}(2009)\citenamefont {Hicken}, \citenamefont {Wood-Vasey}, \citenamefont {Blondin}, \citenamefont {Challis}, \citenamefont {Jha}, \citenamefont {Kelly}, \citenamefont {Rest},\ and\ \citenamefont {Kirshner}}]{hicken2009improved}%
  \BibitemOpen
  \bibfield  {author} {\bibinfo {author} {\bibfnamefont {M.}~\bibnamefont {Hicken}}, \bibinfo {author} {\bibfnamefont {W.~M.}\ \bibnamefont {Wood-Vasey}}, \bibinfo {author} {\bibfnamefont {S.}~\bibnamefont {Blondin}}, \bibinfo {author} {\bibfnamefont {P.}~\bibnamefont {Challis}}, \bibinfo {author} {\bibfnamefont {S.}~\bibnamefont {Jha}}, \bibinfo {author} {\bibfnamefont {P.~L.}\ \bibnamefont {Kelly}}, \bibinfo {author} {\bibfnamefont {A.}~\bibnamefont {Rest}},\ and\ \bibinfo {author} {\bibfnamefont {R.~P.}\ \bibnamefont {Kirshner}},\ }\bibfield  {title} {\bibinfo {title} {Improved dark energy constraints from~ 100 new cfa supernova type ia light curves},\ }\href@noop {} {\bibfield  {journal} {\bibinfo  {journal} {The Astrophysical Journal}\ }\textbf {\bibinfo {volume} {700}},\ \bibinfo {pages} {1097} (\bibinfo {year} {2009})}\BibitemShut {NoStop}%
\bibitem [{\citenamefont {Sullivan}\ \emph {et~al.}(2010)\citenamefont {Sullivan}, \citenamefont {Conley}, \citenamefont {Howell}, \citenamefont {Neill}, \citenamefont {Astier}, \citenamefont {Balland}, \citenamefont {Basa}, \citenamefont {Carlberg}, \citenamefont {Fouchez}, \citenamefont {Guy} \emph {et~al.}}]{sullivan2010dependence}%
  \BibitemOpen
  \bibfield  {author} {\bibinfo {author} {\bibfnamefont {M.}~\bibnamefont {Sullivan}}, \bibinfo {author} {\bibfnamefont {A.}~\bibnamefont {Conley}}, \bibinfo {author} {\bibfnamefont {D.}~\bibnamefont {Howell}}, \bibinfo {author} {\bibfnamefont {J.}~\bibnamefont {Neill}}, \bibinfo {author} {\bibfnamefont {P.}~\bibnamefont {Astier}}, \bibinfo {author} {\bibfnamefont {C.}~\bibnamefont {Balland}}, \bibinfo {author} {\bibfnamefont {S.}~\bibnamefont {Basa}}, \bibinfo {author} {\bibfnamefont {R.}~\bibnamefont {Carlberg}}, \bibinfo {author} {\bibfnamefont {D.}~\bibnamefont {Fouchez}}, \bibinfo {author} {\bibfnamefont {J.}~\bibnamefont {Guy}}, \emph {et~al.},\ }\bibfield  {title} {\bibinfo {title} {The dependence of type ia supernovae luminosities on their host galaxies},\ }\href@noop {} {\bibfield  {journal} {\bibinfo  {journal} {Monthly Notices of the Royal Astronomical Society}\ }\textbf {\bibinfo {volume} {406}},\ \bibinfo {pages} {782} (\bibinfo {year} {2010})}\BibitemShut {NoStop}%
\bibitem [{\citenamefont {Rigault}\ \emph {et~al.}(2013)\citenamefont {Rigault}, \citenamefont {Copin}, \citenamefont {Aldering}, \citenamefont {Antilogus}, \citenamefont {Aragon}, \citenamefont {Bailey}, \citenamefont {Baltay}, \citenamefont {Bongard}, \citenamefont {Buton}, \citenamefont {Canto} \emph {et~al.}}]{rigault2013evidence}%
  \BibitemOpen
  \bibfield  {author} {\bibinfo {author} {\bibfnamefont {M.}~\bibnamefont {Rigault}}, \bibinfo {author} {\bibfnamefont {Y.}~\bibnamefont {Copin}}, \bibinfo {author} {\bibfnamefont {G.}~\bibnamefont {Aldering}}, \bibinfo {author} {\bibfnamefont {P.}~\bibnamefont {Antilogus}}, \bibinfo {author} {\bibfnamefont {C.}~\bibnamefont {Aragon}}, \bibinfo {author} {\bibfnamefont {S.}~\bibnamefont {Bailey}}, \bibinfo {author} {\bibfnamefont {C.}~\bibnamefont {Baltay}}, \bibinfo {author} {\bibfnamefont {S.}~\bibnamefont {Bongard}}, \bibinfo {author} {\bibfnamefont {C.}~\bibnamefont {Buton}}, \bibinfo {author} {\bibfnamefont {A.}~\bibnamefont {Canto}}, \emph {et~al.},\ }\bibfield  {title} {\bibinfo {title} {Evidence of environmental dependencies of type ia supernovae from the nearby supernova factory indicated by local h$\alpha$},\ }\href@noop {} {\bibfield  {journal} {\bibinfo  {journal} {Astronomy \& Astrophysics}\ }\textbf {\bibinfo {volume} {560}},\ \bibinfo {pages} {A66} (\bibinfo {year} {2013})}\BibitemShut {NoStop}%
\bibitem [{\citenamefont {Rigault}\ \emph {et~al.}(2015)\citenamefont {Rigault}, \citenamefont {Aldering}, \citenamefont {Kowalski}, \citenamefont {Copin}, \citenamefont {Antilogus}, \citenamefont {Aragon}, \citenamefont {Bailey}, \citenamefont {Baltay}, \citenamefont {Baugh}, \citenamefont {Bongard} \emph {et~al.}}]{rigault2015confirmation}%
  \BibitemOpen
  \bibfield  {author} {\bibinfo {author} {\bibfnamefont {M.}~\bibnamefont {Rigault}}, \bibinfo {author} {\bibfnamefont {G.}~\bibnamefont {Aldering}}, \bibinfo {author} {\bibfnamefont {M.}~\bibnamefont {Kowalski}}, \bibinfo {author} {\bibfnamefont {Y.}~\bibnamefont {Copin}}, \bibinfo {author} {\bibfnamefont {P.}~\bibnamefont {Antilogus}}, \bibinfo {author} {\bibfnamefont {C.}~\bibnamefont {Aragon}}, \bibinfo {author} {\bibfnamefont {S.}~\bibnamefont {Bailey}}, \bibinfo {author} {\bibfnamefont {C.}~\bibnamefont {Baltay}}, \bibinfo {author} {\bibfnamefont {D.}~\bibnamefont {Baugh}}, \bibinfo {author} {\bibfnamefont {S.}~\bibnamefont {Bongard}}, \emph {et~al.},\ }\bibfield  {title} {\bibinfo {title} {Confirmation of a star formation bias in type ia supernova distances and its effect on the measurement of the hubble constant},\ }\href@noop {} {\bibfield  {journal} {\bibinfo  {journal} {The Astrophysical Journal}\ }\textbf {\bibinfo {volume} {802}},\ \bibinfo {pages} {20} (\bibinfo {year} {2015})}\BibitemShut
  {NoStop}%
\bibitem [{\citenamefont {Choi}\ \emph {et~al.}(2014)\citenamefont {Choi}, \citenamefont {Conroy}, \citenamefont {Moustakas}, \citenamefont {Graves}, \citenamefont {Holden}, \citenamefont {Brodwin}, \citenamefont {Brown},\ and\ \citenamefont {Van~Dokkum}}]{choi2014assembly}%
  \BibitemOpen
  \bibfield  {author} {\bibinfo {author} {\bibfnamefont {J.}~\bibnamefont {Choi}}, \bibinfo {author} {\bibfnamefont {C.}~\bibnamefont {Conroy}}, \bibinfo {author} {\bibfnamefont {J.}~\bibnamefont {Moustakas}}, \bibinfo {author} {\bibfnamefont {G.~J.}\ \bibnamefont {Graves}}, \bibinfo {author} {\bibfnamefont {B.~P.}\ \bibnamefont {Holden}}, \bibinfo {author} {\bibfnamefont {M.}~\bibnamefont {Brodwin}}, \bibinfo {author} {\bibfnamefont {M.~J.}\ \bibnamefont {Brown}},\ and\ \bibinfo {author} {\bibfnamefont {P.~G.}\ \bibnamefont {Van~Dokkum}},\ }\bibfield  {title} {\bibinfo {title} {The assembly histories of quiescent galaxies since z= 0.7 from absorption line spectroscopy},\ }\href@noop {} {\bibfield  {journal} {\bibinfo  {journal} {The Astrophysical Journal}\ }\textbf {\bibinfo {volume} {792}},\ \bibinfo {pages} {95} (\bibinfo {year} {2014})}\BibitemShut {NoStop}%
\bibitem [{\citenamefont {Fumagalli}\ \emph {et~al.}(2016)\citenamefont {Fumagalli}, \citenamefont {Franx}, \citenamefont {van Dokkum}, \citenamefont {Whitaker}, \citenamefont {Skelton}, \citenamefont {Brammer}, \citenamefont {Nelson}, \citenamefont {Maseda}, \citenamefont {Momcheva}, \citenamefont {Kriek} \emph {et~al.}}]{fumagalli2016ages}%
  \BibitemOpen
  \bibfield  {author} {\bibinfo {author} {\bibfnamefont {M.}~\bibnamefont {Fumagalli}}, \bibinfo {author} {\bibfnamefont {M.}~\bibnamefont {Franx}}, \bibinfo {author} {\bibfnamefont {P.}~\bibnamefont {van Dokkum}}, \bibinfo {author} {\bibfnamefont {K.~E.}\ \bibnamefont {Whitaker}}, \bibinfo {author} {\bibfnamefont {R.~E.}\ \bibnamefont {Skelton}}, \bibinfo {author} {\bibfnamefont {G.}~\bibnamefont {Brammer}}, \bibinfo {author} {\bibfnamefont {E.}~\bibnamefont {Nelson}}, \bibinfo {author} {\bibfnamefont {M.}~\bibnamefont {Maseda}}, \bibinfo {author} {\bibfnamefont {I.}~\bibnamefont {Momcheva}}, \bibinfo {author} {\bibfnamefont {M.}~\bibnamefont {Kriek}}, \emph {et~al.},\ }\bibfield  {title} {\bibinfo {title} {Ages of massive galaxies at 0.5> z> 2.0 from 3d-hst rest-frame optical spectroscopy},\ }\href@noop {} {\bibfield  {journal} {\bibinfo  {journal} {The Astrophysical Journal}\ }\textbf {\bibinfo {volume} {822}},\ \bibinfo {pages} {1} (\bibinfo {year} {2016})}\BibitemShut {NoStop}%
\bibitem [{\citenamefont {Uddin}\ \emph {et~al.}(2020)\citenamefont {Uddin}, \citenamefont {Burns}, \citenamefont {Phillips}, \citenamefont {Suntzeff}, \citenamefont {Contreras}, \citenamefont {Hsiao}, \citenamefont {Morrell}, \citenamefont {Galbany}, \citenamefont {Stritzinger}, \citenamefont {Hoeflich} \emph {et~al.}}]{uddin2020carnegie}%
  \BibitemOpen
  \bibfield  {author} {\bibinfo {author} {\bibfnamefont {S.~A.}\ \bibnamefont {Uddin}}, \bibinfo {author} {\bibfnamefont {C.~R.}\ \bibnamefont {Burns}}, \bibinfo {author} {\bibfnamefont {M.}~\bibnamefont {Phillips}}, \bibinfo {author} {\bibfnamefont {N.~B.}\ \bibnamefont {Suntzeff}}, \bibinfo {author} {\bibfnamefont {C.}~\bibnamefont {Contreras}}, \bibinfo {author} {\bibfnamefont {E.~Y.}\ \bibnamefont {Hsiao}}, \bibinfo {author} {\bibfnamefont {N.}~\bibnamefont {Morrell}}, \bibinfo {author} {\bibfnamefont {L.}~\bibnamefont {Galbany}}, \bibinfo {author} {\bibfnamefont {M.}~\bibnamefont {Stritzinger}}, \bibinfo {author} {\bibfnamefont {P.}~\bibnamefont {Hoeflich}}, \emph {et~al.},\ }\bibfield  {title} {\bibinfo {title} {The carnegie supernova project-i: Correlation between type ia supernovae and their host galaxies from optical to near-infrared bands},\ }\href@noop {} {\bibfield  {journal} {\bibinfo  {journal} {The Astrophysical Journal}\ }\textbf {\bibinfo {volume} {901}},\ \bibinfo {pages} {143} (\bibinfo
  {year} {2020})}\BibitemShut {NoStop}%
\bibitem [{\citenamefont {Kim}\ \emph {et~al.}(2019)\citenamefont {Kim}, \citenamefont {KangYijung},\ and\ \citenamefont {LeeYoung-Wook}}]{10.5303/JKAS.2019.52.5.181}%
  \BibitemOpen
  \bibfield  {author} {\bibinfo {author} {\bibfnamefont {Y.-L.}\ \bibnamefont {Kim}}, \bibinfo {author} {\bibnamefont {KangYijung}},\ and\ \bibinfo {author} {\bibnamefont {LeeYoung-Wook}},\ }\bibfield  {title} {\bibinfo {title} {Environmental dependence of type ia supernova luminosities from the yonsei supernova catalog},\ }\href@noop {} {\bibfield  {journal} {\bibinfo  {journal} {Journal of The Korean Astronomical Society}\ }\textbf {\bibinfo {volume} {52}},\ \bibinfo {pages} {181} (\bibinfo {year} {2019})}\BibitemShut {NoStop}%
\bibitem [{\citenamefont {Briday}\ \emph {et~al.}(2022)\citenamefont {Briday}, \citenamefont {Rigault}, \citenamefont {Graziani}, \citenamefont {Copin}, \citenamefont {Aldering}, \citenamefont {Amenouche}, \citenamefont {Brinnel}, \citenamefont {Kim}, \citenamefont {Kim}, \citenamefont {Lezmy} \emph {et~al.}}]{briday2022accuracy}%
  \BibitemOpen
  \bibfield  {author} {\bibinfo {author} {\bibfnamefont {M.}~\bibnamefont {Briday}}, \bibinfo {author} {\bibfnamefont {M.}~\bibnamefont {Rigault}}, \bibinfo {author} {\bibfnamefont {R.}~\bibnamefont {Graziani}}, \bibinfo {author} {\bibfnamefont {Y.}~\bibnamefont {Copin}}, \bibinfo {author} {\bibfnamefont {G.}~\bibnamefont {Aldering}}, \bibinfo {author} {\bibfnamefont {M.}~\bibnamefont {Amenouche}}, \bibinfo {author} {\bibfnamefont {V.}~\bibnamefont {Brinnel}}, \bibinfo {author} {\bibfnamefont {A.}~\bibnamefont {Kim}}, \bibinfo {author} {\bibfnamefont {Y.-L.}\ \bibnamefont {Kim}}, \bibinfo {author} {\bibfnamefont {J.}~\bibnamefont {Lezmy}}, \emph {et~al.},\ }\bibfield  {title} {\bibinfo {title} {Accuracy of environmental tracers and consequences for determining the type ia supernova magnitude step},\ }\href@noop {} {\bibfield  {journal} {\bibinfo  {journal} {Astronomy \& Astrophysics}\ }\textbf {\bibinfo {volume} {657}},\ \bibinfo {pages} {A22} (\bibinfo {year} {2022})}\BibitemShut {NoStop}%
\bibitem [{\citenamefont {Gupta}\ \emph {et~al.}(2011)\citenamefont {Gupta}, \citenamefont {D’Andrea}, \citenamefont {Sako}, \citenamefont {Conroy}, \citenamefont {Smith}, \citenamefont {Bassett}, \citenamefont {Frieman}, \citenamefont {Garnavich}, \citenamefont {Jha}, \citenamefont {Kessler} \emph {et~al.}}]{gupta2011improved}%
  \BibitemOpen
  \bibfield  {author} {\bibinfo {author} {\bibfnamefont {R.~R.}\ \bibnamefont {Gupta}}, \bibinfo {author} {\bibfnamefont {C.~B.}\ \bibnamefont {D’Andrea}}, \bibinfo {author} {\bibfnamefont {M.}~\bibnamefont {Sako}}, \bibinfo {author} {\bibfnamefont {C.}~\bibnamefont {Conroy}}, \bibinfo {author} {\bibfnamefont {M.}~\bibnamefont {Smith}}, \bibinfo {author} {\bibfnamefont {B.}~\bibnamefont {Bassett}}, \bibinfo {author} {\bibfnamefont {J.~A.}\ \bibnamefont {Frieman}}, \bibinfo {author} {\bibfnamefont {P.~M.}\ \bibnamefont {Garnavich}}, \bibinfo {author} {\bibfnamefont {S.~W.}\ \bibnamefont {Jha}}, \bibinfo {author} {\bibfnamefont {R.}~\bibnamefont {Kessler}}, \emph {et~al.},\ }\bibfield  {title} {\bibinfo {title} {Improved constraints on type ia supernova host galaxy properties using multi-wavelength photometry and their correlations with supernova properties},\ }\href@noop {} {\bibfield  {journal} {\bibinfo  {journal} {The Astrophysical Journal}\ }\textbf {\bibinfo {volume} {740}},\ \bibinfo {pages} {92}
  (\bibinfo {year} {2011})}\BibitemShut {NoStop}%
\bibitem [{\citenamefont {Kang}\ \emph {et~al.}(2016)\citenamefont {Kang}, \citenamefont {Kim}, \citenamefont {Lim}, \citenamefont {Chung},\ and\ \citenamefont {Lee}}]{kang2016early}%
  \BibitemOpen
  \bibfield  {author} {\bibinfo {author} {\bibfnamefont {Y.}~\bibnamefont {Kang}}, \bibinfo {author} {\bibfnamefont {Y.-L.}\ \bibnamefont {Kim}}, \bibinfo {author} {\bibfnamefont {D.}~\bibnamefont {Lim}}, \bibinfo {author} {\bibfnamefont {C.}~\bibnamefont {Chung}},\ and\ \bibinfo {author} {\bibfnamefont {Y.-W.}\ \bibnamefont {Lee}},\ }\bibfield  {title} {\bibinfo {title} {Early-type host galaxies of type ia supernovae. i. evidence for downsizing},\ }\href@noop {} {\bibfield  {journal} {\bibinfo  {journal} {The Astrophysical Journal Supplement Series}\ }\textbf {\bibinfo {volume} {223}},\ \bibinfo {pages} {7} (\bibinfo {year} {2016})}\BibitemShut {NoStop}%
\bibitem [{\citenamefont {Rose}\ \emph {et~al.}(2019)\citenamefont {Rose}, \citenamefont {Garnavich},\ and\ \citenamefont {Berg}}]{rose2019think}%
  \BibitemOpen
  \bibfield  {author} {\bibinfo {author} {\bibfnamefont {B.}~\bibnamefont {Rose}}, \bibinfo {author} {\bibfnamefont {P.}~\bibnamefont {Garnavich}},\ and\ \bibinfo {author} {\bibfnamefont {M.}~\bibnamefont {Berg}},\ }\bibfield  {title} {\bibinfo {title} {Think global, act local: the influence of environment age and host mass on type ia supernova light curves},\ }\href@noop {} {\bibfield  {journal} {\bibinfo  {journal} {The Astrophysical Journal}\ }\textbf {\bibinfo {volume} {874}},\ \bibinfo {pages} {32} (\bibinfo {year} {2019})}\BibitemShut {NoStop}%
\bibitem [{\citenamefont {Rose}\ \emph {et~al.}(2020)\citenamefont {Rose}, \citenamefont {Rubin}, \citenamefont {Cikota}, \citenamefont {Deustua}, \citenamefont {Dixon}, \citenamefont {Fruchter}, \citenamefont {Jones}, \citenamefont {Riess},\ and\ \citenamefont {Scolnic}}]{rose2020evidence}%
  \BibitemOpen
  \bibfield  {author} {\bibinfo {author} {\bibfnamefont {B.}~\bibnamefont {Rose}}, \bibinfo {author} {\bibfnamefont {D.}~\bibnamefont {Rubin}}, \bibinfo {author} {\bibfnamefont {A.}~\bibnamefont {Cikota}}, \bibinfo {author} {\bibfnamefont {S.}~\bibnamefont {Deustua}}, \bibinfo {author} {\bibfnamefont {S.}~\bibnamefont {Dixon}}, \bibinfo {author} {\bibfnamefont {A.}~\bibnamefont {Fruchter}}, \bibinfo {author} {\bibfnamefont {D.}~\bibnamefont {Jones}}, \bibinfo {author} {\bibfnamefont {A.}~\bibnamefont {Riess}},\ and\ \bibinfo {author} {\bibfnamefont {D.}~\bibnamefont {Scolnic}},\ }\bibfield  {title} {\bibinfo {title} {Evidence for cosmic acceleration is robust to observed correlations between type ia supernova luminosity and stellar age},\ }\href@noop {} {\bibfield  {journal} {\bibinfo  {journal} {The Astrophysical Journal Letters}\ }\textbf {\bibinfo {volume} {896}},\ \bibinfo {pages} {L4} (\bibinfo {year} {2020})}\BibitemShut {NoStop}%
\bibitem [{\citenamefont {Kang}\ \emph {et~al.}(2020)\citenamefont {Kang}, \citenamefont {Lee}, \citenamefont {Kim}, \citenamefont {Chung},\ and\ \citenamefont {Ree}}]{kang2020early}%
  \BibitemOpen
  \bibfield  {author} {\bibinfo {author} {\bibfnamefont {Y.}~\bibnamefont {Kang}}, \bibinfo {author} {\bibfnamefont {Y.-W.}\ \bibnamefont {Lee}}, \bibinfo {author} {\bibfnamefont {Y.-L.}\ \bibnamefont {Kim}}, \bibinfo {author} {\bibfnamefont {C.}~\bibnamefont {Chung}},\ and\ \bibinfo {author} {\bibfnamefont {C.~H.}\ \bibnamefont {Ree}},\ }\bibfield  {title} {\bibinfo {title} {Early-type host galaxies of type ia supernovae. ii. evidence for luminosity evolution in supernova cosmology},\ }\href@noop {} {\bibfield  {journal} {\bibinfo  {journal} {The Astrophysical Journal}\ }\textbf {\bibinfo {volume} {889}},\ \bibinfo {pages} {8} (\bibinfo {year} {2020})}\BibitemShut {NoStop}%
\bibitem [{\citenamefont {Lee}\ \emph {et~al.}(2020)\citenamefont {Lee}, \citenamefont {Chung}, \citenamefont {Kang},\ and\ \citenamefont {Jee}}]{lee2020further}%
  \BibitemOpen
  \bibfield  {author} {\bibinfo {author} {\bibfnamefont {Y.-W.}\ \bibnamefont {Lee}}, \bibinfo {author} {\bibfnamefont {C.}~\bibnamefont {Chung}}, \bibinfo {author} {\bibfnamefont {Y.}~\bibnamefont {Kang}},\ and\ \bibinfo {author} {\bibfnamefont {M.~J.}\ \bibnamefont {Jee}},\ }\bibfield  {title} {\bibinfo {title} {Further evidence for significant luminosity evolution in supernova cosmology},\ }\href@noop {} {\bibfield  {journal} {\bibinfo  {journal} {The Astrophysical Journal}\ }\textbf {\bibinfo {volume} {903}},\ \bibinfo {pages} {22} (\bibinfo {year} {2020})}\BibitemShut {NoStop}%
\bibitem [{\citenamefont {Rigault}\ \emph {et~al.}(2020)\citenamefont {Rigault}, \citenamefont {Brinnel}, \citenamefont {Aldering}, \citenamefont {Antilogus}, \citenamefont {Aragon}, \citenamefont {Bailey}, \citenamefont {Baltay}, \citenamefont {Barbary}, \citenamefont {Bongard}, \citenamefont {Boone} \emph {et~al.}}]{rigault2020strong}%
  \BibitemOpen
  \bibfield  {author} {\bibinfo {author} {\bibfnamefont {M.}~\bibnamefont {Rigault}}, \bibinfo {author} {\bibfnamefont {V.}~\bibnamefont {Brinnel}}, \bibinfo {author} {\bibfnamefont {G.}~\bibnamefont {Aldering}}, \bibinfo {author} {\bibfnamefont {P.}~\bibnamefont {Antilogus}}, \bibinfo {author} {\bibfnamefont {C.}~\bibnamefont {Aragon}}, \bibinfo {author} {\bibfnamefont {S.}~\bibnamefont {Bailey}}, \bibinfo {author} {\bibfnamefont {C.}~\bibnamefont {Baltay}}, \bibinfo {author} {\bibfnamefont {K.}~\bibnamefont {Barbary}}, \bibinfo {author} {\bibfnamefont {S.}~\bibnamefont {Bongard}}, \bibinfo {author} {\bibfnamefont {K.}~\bibnamefont {Boone}}, \emph {et~al.},\ }\bibfield  {title} {\bibinfo {title} {Strong dependence of type ia supernova standardization on the local specific star formation rate},\ }\href@noop {} {\bibfield  {journal} {\bibinfo  {journal} {Astronomy \& Astrophysics}\ }\textbf {\bibinfo {volume} {644}},\ \bibinfo {pages} {A176} (\bibinfo {year} {2020})}\BibitemShut {NoStop}%
\bibitem [{\citenamefont {Zhang}\ \emph {et~al.}(2021)\citenamefont {Zhang}, \citenamefont {Murakami}, \citenamefont {Stahl}, \citenamefont {Patra},\ and\ \citenamefont {Filippenko}}]{zhang2021improving}%
  \BibitemOpen
  \bibfield  {author} {\bibinfo {author} {\bibfnamefont {K.~D.}\ \bibnamefont {Zhang}}, \bibinfo {author} {\bibfnamefont {Y.~S.}\ \bibnamefont {Murakami}}, \bibinfo {author} {\bibfnamefont {B.~E.}\ \bibnamefont {Stahl}}, \bibinfo {author} {\bibfnamefont {K.~C.}\ \bibnamefont {Patra}},\ and\ \bibinfo {author} {\bibfnamefont {A.~V.}\ \bibnamefont {Filippenko}},\ }\bibfield  {title} {\bibinfo {title} {Improving bayesian posterior correlation analysis on type ia supernova luminosity evolution},\ }\href@noop {} {\bibfield  {journal} {\bibinfo  {journal} {Monthly Notices of the Royal Astronomical Society: Letters}\ }\textbf {\bibinfo {volume} {503}},\ \bibinfo {pages} {L33} (\bibinfo {year} {2021})}\BibitemShut {NoStop}%
\bibitem [{\citenamefont {Lee}\ \emph {et~al.}(2022)\citenamefont {Lee}, \citenamefont {Chung}, \citenamefont {Demarque}, \citenamefont {Park}, \citenamefont {Son},\ and\ \citenamefont {Kang}}]{lee2022evidence}%
  \BibitemOpen
  \bibfield  {author} {\bibinfo {author} {\bibfnamefont {Y.-W.}\ \bibnamefont {Lee}}, \bibinfo {author} {\bibfnamefont {C.}~\bibnamefont {Chung}}, \bibinfo {author} {\bibfnamefont {P.}~\bibnamefont {Demarque}}, \bibinfo {author} {\bibfnamefont {S.}~\bibnamefont {Park}}, \bibinfo {author} {\bibfnamefont {J.}~\bibnamefont {Son}},\ and\ \bibinfo {author} {\bibfnamefont {Y.}~\bibnamefont {Kang}},\ }\bibfield  {title} {\bibinfo {title} {Evidence for strong progenitor age dependence of type ia supernova luminosity standardization process},\ }\href@noop {} {\bibfield  {journal} {\bibinfo  {journal} {Monthly Notices of the Royal Astronomical Society}\ }\textbf {\bibinfo {volume} {517}},\ \bibinfo {pages} {2697} (\bibinfo {year} {2022})}\BibitemShut {NoStop}%
\bibitem [{\citenamefont {Wiseman}\ \emph {et~al.}(2022)\citenamefont {Wiseman}, \citenamefont {Vincenzi}, \citenamefont {Sullivan}, \citenamefont {Kelsey}, \citenamefont {Popovic}, \citenamefont {Rose}, \citenamefont {Brout}, \citenamefont {Davis}, \citenamefont {Frohmaier}, \citenamefont {Galbany} \emph {et~al.}}]{wiseman2022galaxy}%
  \BibitemOpen
  \bibfield  {author} {\bibinfo {author} {\bibfnamefont {P.}~\bibnamefont {Wiseman}}, \bibinfo {author} {\bibfnamefont {M.}~\bibnamefont {Vincenzi}}, \bibinfo {author} {\bibfnamefont {M.}~\bibnamefont {Sullivan}}, \bibinfo {author} {\bibfnamefont {L.}~\bibnamefont {Kelsey}}, \bibinfo {author} {\bibfnamefont {B.}~\bibnamefont {Popovic}}, \bibinfo {author} {\bibfnamefont {B.}~\bibnamefont {Rose}}, \bibinfo {author} {\bibfnamefont {D.}~\bibnamefont {Brout}}, \bibinfo {author} {\bibfnamefont {T.}~\bibnamefont {Davis}}, \bibinfo {author} {\bibfnamefont {C.}~\bibnamefont {Frohmaier}}, \bibinfo {author} {\bibfnamefont {L.}~\bibnamefont {Galbany}}, \emph {et~al.},\ }\bibfield  {title} {\bibinfo {title} {A galaxy-driven model of type ia supernova luminosity variations},\ }\href@noop {} {\bibfield  {journal} {\bibinfo  {journal} {Monthly Notices of the Royal Astronomical Society}\ }\textbf {\bibinfo {volume} {515}},\ \bibinfo {pages} {4587} (\bibinfo {year} {2022})}\BibitemShut {NoStop}%
\bibitem [{\citenamefont {Wiseman}\ \emph {et~al.}(2023)\citenamefont {Wiseman}, \citenamefont {Sullivan}, \citenamefont {Smith},\ and\ \citenamefont {Popovic}}]{wiseman2023further}%
  \BibitemOpen
  \bibfield  {author} {\bibinfo {author} {\bibfnamefont {P.}~\bibnamefont {Wiseman}}, \bibinfo {author} {\bibfnamefont {M.}~\bibnamefont {Sullivan}}, \bibinfo {author} {\bibfnamefont {M.}~\bibnamefont {Smith}},\ and\ \bibinfo {author} {\bibfnamefont {B.}~\bibnamefont {Popovic}},\ }\bibfield  {title} {\bibinfo {title} {Further evidence that galaxy age drives observed type ia supernova luminosity differences},\ }\href@noop {} {\bibfield  {journal} {\bibinfo  {journal} {Monthly Notices of the Royal Astronomical Society}\ }\textbf {\bibinfo {volume} {520}},\ \bibinfo {pages} {6214} (\bibinfo {year} {2023})}\BibitemShut {NoStop}%
\bibitem [{\citenamefont {Campbell}\ \emph {et~al.}(2013)\citenamefont {Campbell}, \citenamefont {D’Andrea}, \citenamefont {Nichol}, \citenamefont {Sako}, \citenamefont {Smith}, \citenamefont {Lampeitl}, \citenamefont {Olmstead}, \citenamefont {Bassett}, \citenamefont {Biswas}, \citenamefont {Brown} \emph {et~al.}}]{campbell2013cosmology}%
  \BibitemOpen
  \bibfield  {author} {\bibinfo {author} {\bibfnamefont {H.}~\bibnamefont {Campbell}}, \bibinfo {author} {\bibfnamefont {C.~B.}\ \bibnamefont {D’Andrea}}, \bibinfo {author} {\bibfnamefont {R.~C.}\ \bibnamefont {Nichol}}, \bibinfo {author} {\bibfnamefont {M.}~\bibnamefont {Sako}}, \bibinfo {author} {\bibfnamefont {M.}~\bibnamefont {Smith}}, \bibinfo {author} {\bibfnamefont {H.}~\bibnamefont {Lampeitl}}, \bibinfo {author} {\bibfnamefont {M.~D.}\ \bibnamefont {Olmstead}}, \bibinfo {author} {\bibfnamefont {B.}~\bibnamefont {Bassett}}, \bibinfo {author} {\bibfnamefont {R.}~\bibnamefont {Biswas}}, \bibinfo {author} {\bibfnamefont {P.}~\bibnamefont {Brown}}, \emph {et~al.},\ }\bibfield  {title} {\bibinfo {title} {Cosmology with photometrically classified type ia supernovae from the sdss-ii supernova survey},\ }\href@noop {} {\bibfield  {journal} {\bibinfo  {journal} {The Astrophysical Journal}\ }\textbf {\bibinfo {volume} {763}},\ \bibinfo {pages} {88} (\bibinfo {year} {2013})}\BibitemShut {NoStop}%
\bibitem [{\citenamefont {Huang}(2020)}]{huang2020supernova}%
  \BibitemOpen
  \bibfield  {author} {\bibinfo {author} {\bibfnamefont {Z.}~\bibnamefont {Huang}},\ }\bibfield  {title} {\bibinfo {title} {Supernova magnitude evolution and page approximation},\ }\href@noop {} {\bibfield  {journal} {\bibinfo  {journal} {The Astrophysical Journal Letters}\ }\textbf {\bibinfo {volume} {892}},\ \bibinfo {pages} {L28} (\bibinfo {year} {2020})}\BibitemShut {NoStop}%
\bibitem [{\citenamefont {Scolnic}\ \emph {et~al.}(2018)\citenamefont {Scolnic}, \citenamefont {Jones}, \citenamefont {Rest}, \citenamefont {Pan}, \citenamefont {Chornock}, \citenamefont {Foley}, \citenamefont {Huber}, \citenamefont {Kessler}, \citenamefont {Narayan}, \citenamefont {Riess} \emph {et~al.}}]{scolnic2018complete}%
  \BibitemOpen
  \bibfield  {author} {\bibinfo {author} {\bibfnamefont {D.~M.}\ \bibnamefont {Scolnic}}, \bibinfo {author} {\bibfnamefont {D.}~\bibnamefont {Jones}}, \bibinfo {author} {\bibfnamefont {A.}~\bibnamefont {Rest}}, \bibinfo {author} {\bibfnamefont {Y.}~\bibnamefont {Pan}}, \bibinfo {author} {\bibfnamefont {R.}~\bibnamefont {Chornock}}, \bibinfo {author} {\bibfnamefont {R.}~\bibnamefont {Foley}}, \bibinfo {author} {\bibfnamefont {M.}~\bibnamefont {Huber}}, \bibinfo {author} {\bibfnamefont {R.}~\bibnamefont {Kessler}}, \bibinfo {author} {\bibfnamefont {G.}~\bibnamefont {Narayan}}, \bibinfo {author} {\bibfnamefont {A.}~\bibnamefont {Riess}}, \emph {et~al.},\ }\bibfield  {title} {\bibinfo {title} {The complete light-curve sample of spectroscopically confirmed sne ia from pan-starrs1 and cosmological constraints from the combined pantheon sample},\ }\href@noop {} {\bibfield  {journal} {\bibinfo  {journal} {The Astrophysical Journal}\ }\textbf {\bibinfo {volume} {859}},\ \bibinfo {pages} {101} (\bibinfo {year}
  {2018})}\BibitemShut {NoStop}%
\bibitem [{\citenamefont {Guy}\ \emph {et~al.}(2010)\citenamefont {Guy}, \citenamefont {Sullivan}, \citenamefont {Conley}, \citenamefont {Regnault}, \citenamefont {Astier}, \citenamefont {Balland}, \citenamefont {Basa}, \citenamefont {Carlberg}, \citenamefont {Fouchez}, \citenamefont {Hardin} \emph {et~al.}}]{guy2010supernova}%
  \BibitemOpen
  \bibfield  {author} {\bibinfo {author} {\bibfnamefont {J.}~\bibnamefont {Guy}}, \bibinfo {author} {\bibfnamefont {M.}~\bibnamefont {Sullivan}}, \bibinfo {author} {\bibfnamefont {A.}~\bibnamefont {Conley}}, \bibinfo {author} {\bibfnamefont {N.}~\bibnamefont {Regnault}}, \bibinfo {author} {\bibfnamefont {P.}~\bibnamefont {Astier}}, \bibinfo {author} {\bibfnamefont {C.}~\bibnamefont {Balland}}, \bibinfo {author} {\bibfnamefont {S.}~\bibnamefont {Basa}}, \bibinfo {author} {\bibfnamefont {R.}~\bibnamefont {Carlberg}}, \bibinfo {author} {\bibfnamefont {D.}~\bibnamefont {Fouchez}}, \bibinfo {author} {\bibfnamefont {D.}~\bibnamefont {Hardin}}, \emph {et~al.},\ }\bibfield  {title} {\bibinfo {title} {The supernova legacy survey 3-year sample: Type ia supernovae photometric distances and cosmological constraints},\ }\href@noop {} {\bibfield  {journal} {\bibinfo  {journal} {Astronomy \& Astrophysics}\ }\textbf {\bibinfo {volume} {523}},\ \bibinfo {pages} {A7} (\bibinfo {year} {2010})}\BibitemShut {NoStop}%
\bibitem [{\citenamefont {Conley}\ \emph {et~al.}(2010)\citenamefont {Conley}, \citenamefont {Guy}, \citenamefont {Sullivan}, \citenamefont {Regnault}, \citenamefont {Astier}, \citenamefont {Balland}, \citenamefont {Basa}, \citenamefont {Carlberg}, \citenamefont {Fouchez}, \citenamefont {Hardin} \emph {et~al.}}]{conley2010supernova}%
  \BibitemOpen
  \bibfield  {author} {\bibinfo {author} {\bibfnamefont {A.}~\bibnamefont {Conley}}, \bibinfo {author} {\bibfnamefont {J.}~\bibnamefont {Guy}}, \bibinfo {author} {\bibfnamefont {M.}~\bibnamefont {Sullivan}}, \bibinfo {author} {\bibfnamefont {N.}~\bibnamefont {Regnault}}, \bibinfo {author} {\bibfnamefont {P.}~\bibnamefont {Astier}}, \bibinfo {author} {\bibfnamefont {C.}~\bibnamefont {Balland}}, \bibinfo {author} {\bibfnamefont {S.}~\bibnamefont {Basa}}, \bibinfo {author} {\bibfnamefont {R.}~\bibnamefont {Carlberg}}, \bibinfo {author} {\bibfnamefont {D.}~\bibnamefont {Fouchez}}, \bibinfo {author} {\bibfnamefont {D.}~\bibnamefont {Hardin}}, \emph {et~al.},\ }\bibfield  {title} {\bibinfo {title} {Supernova constraints and systematic uncertainties from the first three years of the supernova legacy survey},\ }\href@noop {} {\bibfield  {journal} {\bibinfo  {journal} {The Astrophysical Journal Supplement Series}\ }\textbf {\bibinfo {volume} {192}},\ \bibinfo {pages} {1} (\bibinfo {year} {2010})}\BibitemShut {NoStop}%
\bibitem [{\citenamefont {Aghanim}\ \emph {et~al.}(2020)\citenamefont {Aghanim}, \citenamefont {Akrami}, \citenamefont {Ashdown}, \citenamefont {Aumont}, \citenamefont {Baccigalupi}, \citenamefont {Ballardini}, \citenamefont {Banday}, \citenamefont {Barreiro}, \citenamefont {Bartolo}, \citenamefont {Basak} \emph {et~al.}}]{Planck18}%
  \BibitemOpen
  \bibfield  {author} {\bibinfo {author} {\bibfnamefont {N.}~\bibnamefont {Aghanim}}, \bibinfo {author} {\bibfnamefont {Y.}~\bibnamefont {Akrami}}, \bibinfo {author} {\bibfnamefont {M.}~\bibnamefont {Ashdown}}, \bibinfo {author} {\bibfnamefont {J.}~\bibnamefont {Aumont}}, \bibinfo {author} {\bibfnamefont {C.}~\bibnamefont {Baccigalupi}}, \bibinfo {author} {\bibfnamefont {M.}~\bibnamefont {Ballardini}}, \bibinfo {author} {\bibfnamefont {A.}~\bibnamefont {Banday}}, \bibinfo {author} {\bibfnamefont {R.}~\bibnamefont {Barreiro}}, \bibinfo {author} {\bibfnamefont {N.}~\bibnamefont {Bartolo}}, \bibinfo {author} {\bibfnamefont {S.}~\bibnamefont {Basak}}, \emph {et~al.},\ }\bibfield  {title} {\bibinfo {title} {Planck 2018 results-vi. cosmological parameters},\ }\href@noop {} {\bibfield  {journal} {\bibinfo  {journal} {Astronomy \& Astrophysics}\ }\textbf {\bibinfo {volume} {641}},\ \bibinfo {pages} {A6} (\bibinfo {year} {2020})}\BibitemShut {NoStop}%
\bibitem [{Note1()}]{Note1}%
  \BibitemOpen
  \bibinfo {note} {The matter abundance $\Omega _m$ here is constrained by the Pantheon SN Ia data set, while $\Omega _m$ in Table~\ref {iii} is constrained by the 102 low-redshift SN Ia samples, so they are different.}\BibitemShut {Stop}%
\bibitem [{\citenamefont {Luo}\ \emph {et~al.}(2020)\citenamefont {Luo}, \citenamefont {Huang}, \citenamefont {Qian},\ and\ \citenamefont {Huang}}]{luo2020reaffirming}%
  \BibitemOpen
  \bibfield  {author} {\bibinfo {author} {\bibfnamefont {X.}~\bibnamefont {Luo}}, \bibinfo {author} {\bibfnamefont {Z.}~\bibnamefont {Huang}}, \bibinfo {author} {\bibfnamefont {Q.}~\bibnamefont {Qian}},\ and\ \bibinfo {author} {\bibfnamefont {L.}~\bibnamefont {Huang}},\ }\bibfield  {title} {\bibinfo {title} {Reaffirming the cosmic acceleration without supernovae and the cosmic microwave background},\ }\href@noop {} {\bibfield  {journal} {\bibinfo  {journal} {The Astrophysical Journal}\ }\textbf {\bibinfo {volume} {905}},\ \bibinfo {pages} {53} (\bibinfo {year} {2020})}\BibitemShut {NoStop}%
\bibitem [{\citenamefont {Huang}\ \emph {et~al.}(2021{\natexlab{a}})\citenamefont {Huang}, \citenamefont {Huang}, \citenamefont {Li},\ and\ \citenamefont {Zhou}}]{MAPAge}%
  \BibitemOpen
  \bibfield  {author} {\bibinfo {author} {\bibfnamefont {L.}~\bibnamefont {Huang}}, \bibinfo {author} {\bibfnamefont {Z.}~\bibnamefont {Huang}}, \bibinfo {author} {\bibfnamefont {Z.}~\bibnamefont {Li}},\ and\ \bibinfo {author} {\bibfnamefont {H.}~\bibnamefont {Zhou}},\ }\bibfield  {title} {\bibinfo {title} {{A More Accurate Parameterization based on cosmic Age (MAPAge)}},\ }\href@noop {} {\bibfield  {journal} {\bibinfo  {journal} {{RAA}}\ }\textbf {\bibinfo {volume} {21}},\ \bibinfo {pages} {277} (\bibinfo {year} {2021}{\natexlab{a}})}\BibitemShut {NoStop}%
\bibitem [{\citenamefont {Huang}\ \emph {et~al.}(2021{\natexlab{b}})\citenamefont {Huang}, \citenamefont {Zhiqi}, \citenamefont {Luo},\ and\ \citenamefont {Yuhong}}]{GRB2021PAge}%
  \BibitemOpen
  \bibfield  {author} {\bibinfo {author} {\bibfnamefont {L.}~\bibnamefont {Huang}}, \bibinfo {author} {\bibfnamefont {H.}~\bibnamefont {Zhiqi}}, \bibinfo {author} {\bibfnamefont {X.}~\bibnamefont {Luo}},\ and\ \bibinfo {author} {\bibfnamefont {F.}~\bibnamefont {Yuhong}},\ }\bibfield  {title} {\bibinfo {title} {{Reconciling low and high redshift GRB luminosity correlations}},\ }\href@noop {} {\bibfield  {journal} {\bibinfo  {journal} {{PRD}}\ }\textbf {\bibinfo {volume} {103}},\ \bibinfo {pages} {123521} (\bibinfo {year} {2021}{\natexlab{b}})}\BibitemShut {NoStop}%
\bibitem [{\citenamefont {Huang}\ \emph {et~al.}(2022)\citenamefont {Huang}, \citenamefont {Huang}, \citenamefont {Zhou},\ and\ \citenamefont {Li}}]{huang2022s}%
  \BibitemOpen
  \bibfield  {author} {\bibinfo {author} {\bibfnamefont {L.}~\bibnamefont {Huang}}, \bibinfo {author} {\bibfnamefont {Z.}~\bibnamefont {Huang}}, \bibinfo {author} {\bibfnamefont {H.}~\bibnamefont {Zhou}},\ and\ \bibinfo {author} {\bibfnamefont {Z.}~\bibnamefont {Li}},\ }\bibfield  {title} {\bibinfo {title} {The s 8 tension in light of updated redshift-space distortion data and page approximation},\ }\href@noop {} {\bibfield  {journal} {\bibinfo  {journal} {Science China Physics, Mechanics \& Astronomy}\ }\textbf {\bibinfo {volume} {65}},\ \bibinfo {pages} {239512} (\bibinfo {year} {2022})}\BibitemShut {NoStop}%
\bibitem [{\citenamefont {Huang}(2022)}]{Huang2022PAge}%
  \BibitemOpen
  \bibfield  {author} {\bibinfo {author} {\bibfnamefont {Z.}~\bibnamefont {Huang}},\ }\bibfield  {title} {\bibinfo {title} {{Thawing k-essence dark energy in the PAge space}},\ }\href@noop {} {\bibfield  {journal} {\bibinfo  {journal} {{Communications in Theoretical Physics}}\ }\textbf {\bibinfo {volume} {74}},\ \bibinfo {pages} {095404} (\bibinfo {year} {2022})}\BibitemShut {NoStop}%
\bibitem [{\citenamefont {Cai}\ \emph {et~al.}(2022{\natexlab{a}})\citenamefont {Cai}, \citenamefont {Guo}, \citenamefont {Wang}, \citenamefont {Yu},\ and\ \citenamefont {Zhou}}]{cai2022no2}%
  \BibitemOpen
  \bibfield  {author} {\bibinfo {author} {\bibfnamefont {R.-G.}\ \bibnamefont {Cai}}, \bibinfo {author} {\bibfnamefont {Z.-K.}\ \bibnamefont {Guo}}, \bibinfo {author} {\bibfnamefont {S.-J.}\ \bibnamefont {Wang}}, \bibinfo {author} {\bibfnamefont {W.-W.}\ \bibnamefont {Yu}},\ and\ \bibinfo {author} {\bibfnamefont {Y.}~\bibnamefont {Zhou}},\ }\bibfield  {title} {\bibinfo {title} {No-go guide for the hubble tension: matter perturbations},\ }\href@noop {} {\bibfield  {journal} {\bibinfo  {journal} {arXiv preprint arXiv:2202.12214}\ } (\bibinfo {year} {2022}{\natexlab{a}})}\BibitemShut {NoStop}%
\bibitem [{\citenamefont {Cai}\ \emph {et~al.}(2022{\natexlab{b}})\citenamefont {Cai}, \citenamefont {Guo}, \citenamefont {Wang}, \citenamefont {Yu},\ and\ \citenamefont {Zhou}}]{cai2022no1}%
  \BibitemOpen
  \bibfield  {author} {\bibinfo {author} {\bibfnamefont {R.-G.}\ \bibnamefont {Cai}}, \bibinfo {author} {\bibfnamefont {Z.-K.}\ \bibnamefont {Guo}}, \bibinfo {author} {\bibfnamefont {S.-J.}\ \bibnamefont {Wang}}, \bibinfo {author} {\bibfnamefont {W.-W.}\ \bibnamefont {Yu}},\ and\ \bibinfo {author} {\bibfnamefont {Y.}~\bibnamefont {Zhou}},\ }\bibfield  {title} {\bibinfo {title} {No-go guide for the hubble tension: Late-time solutions},\ }\href@noop {} {\bibfield  {journal} {\bibinfo  {journal} {Physical Review D}\ }\textbf {\bibinfo {volume} {105}},\ \bibinfo {pages} {L021301} (\bibinfo {year} {2022}{\natexlab{b}})}\BibitemShut {NoStop}%
\bibitem [{\citenamefont {Li}\ \emph {et~al.}(2022)\citenamefont {Li}, \citenamefont {Huang},\ and\ \citenamefont {Wang}}]{li2022redshift}%
  \BibitemOpen
  \bibfield  {author} {\bibinfo {author} {\bibfnamefont {Z.}~\bibnamefont {Li}}, \bibinfo {author} {\bibfnamefont {L.}~\bibnamefont {Huang}},\ and\ \bibinfo {author} {\bibfnamefont {J.}~\bibnamefont {Wang}},\ }\bibfield  {title} {\bibinfo {title} {Redshift evolution and non-universal dispersion of quasar luminosity correlation},\ }\href@noop {} {\bibfield  {journal} {\bibinfo  {journal} {Monthly Notices of the Royal Astronomical Society}\ }\textbf {\bibinfo {volume} {517}},\ \bibinfo {pages} {1901} (\bibinfo {year} {2022})}\BibitemShut {NoStop}%
\bibitem [{Note2()}]{Note2}%
  \BibitemOpen
  \bibinfo {note} {Https://svn.sdss.org/public/data/eboss/DR16cosmo/\\ tags/v1\protect \_0\protect \_0/likelihoods /BAO-only/}\BibitemShut {NoStop}%
\bibitem [{\citenamefont {Alam}\ \emph {et~al.}(2021)\citenamefont {Alam}, \citenamefont {Aubert}, \citenamefont {Avila}, \citenamefont {Balland}, \citenamefont {Bautista}, \citenamefont {Bershady}, \citenamefont {Bizyaev}, \citenamefont {Blanton}, \citenamefont {Bolton}, \citenamefont {Bovy} \emph {et~al.}}]{alam2021completed}%
  \BibitemOpen
  \bibfield  {author} {\bibinfo {author} {\bibfnamefont {S.}~\bibnamefont {Alam}}, \bibinfo {author} {\bibfnamefont {M.}~\bibnamefont {Aubert}}, \bibinfo {author} {\bibfnamefont {S.}~\bibnamefont {Avila}}, \bibinfo {author} {\bibfnamefont {C.}~\bibnamefont {Balland}}, \bibinfo {author} {\bibfnamefont {J.~E.}\ \bibnamefont {Bautista}}, \bibinfo {author} {\bibfnamefont {M.~A.}\ \bibnamefont {Bershady}}, \bibinfo {author} {\bibfnamefont {D.}~\bibnamefont {Bizyaev}}, \bibinfo {author} {\bibfnamefont {M.~R.}\ \bibnamefont {Blanton}}, \bibinfo {author} {\bibfnamefont {A.~S.}\ \bibnamefont {Bolton}}, \bibinfo {author} {\bibfnamefont {J.}~\bibnamefont {Bovy}}, \emph {et~al.},\ }\bibfield  {title} {\bibinfo {title} {Completed sdss-iv extended baryon oscillation spectroscopic survey: Cosmological implications from two decades of spectroscopic surveys at the apache point observatory},\ }\href@noop {} {\bibfield  {journal} {\bibinfo  {journal} {Physical Review D}\ }\textbf {\bibinfo {volume} {103}},\ \bibinfo {pages}
  {083533} (\bibinfo {year} {2021})}\BibitemShut {NoStop}%
\bibitem [{\citenamefont {Beutler}\ \emph {et~al.}(2011)\citenamefont {Beutler}, \citenamefont {Blake}, \citenamefont {Colless}, \citenamefont {Jones}, \citenamefont {Staveley-Smith}, \citenamefont {Campbell}, \citenamefont {Parker}, \citenamefont {Saunders},\ and\ \citenamefont {Watson}}]{beutler20116df}%
  \BibitemOpen
  \bibfield  {author} {\bibinfo {author} {\bibfnamefont {F.}~\bibnamefont {Beutler}}, \bibinfo {author} {\bibfnamefont {C.}~\bibnamefont {Blake}}, \bibinfo {author} {\bibfnamefont {M.}~\bibnamefont {Colless}}, \bibinfo {author} {\bibfnamefont {D.~H.}\ \bibnamefont {Jones}}, \bibinfo {author} {\bibfnamefont {L.}~\bibnamefont {Staveley-Smith}}, \bibinfo {author} {\bibfnamefont {L.}~\bibnamefont {Campbell}}, \bibinfo {author} {\bibfnamefont {Q.}~\bibnamefont {Parker}}, \bibinfo {author} {\bibfnamefont {W.}~\bibnamefont {Saunders}},\ and\ \bibinfo {author} {\bibfnamefont {F.}~\bibnamefont {Watson}},\ }\bibfield  {title} {\bibinfo {title} {The 6df galaxy survey: baryon acoustic oscillations and the local hubble constant},\ }\href@noop {} {\bibfield  {journal} {\bibinfo  {journal} {Monthly Notices of the Royal Astronomical Society}\ }\textbf {\bibinfo {volume} {416}},\ \bibinfo {pages} {3017} (\bibinfo {year} {2011})}\BibitemShut {NoStop}%
\bibitem [{\citenamefont {Abbott}\ \emph {et~al.}(2022)\citenamefont {Abbott}, \citenamefont {Aguena}, \citenamefont {Allam}, \citenamefont {Amon}, \citenamefont {Andrade-Oliveira}, \citenamefont {Asorey}, \citenamefont {Avila}, \citenamefont {Bernstein}, \citenamefont {Bertin}, \citenamefont {Brandao-Souza} \emph {et~al.}}]{abbott2022dark}%
  \BibitemOpen
  \bibfield  {author} {\bibinfo {author} {\bibfnamefont {T.}~\bibnamefont {Abbott}}, \bibinfo {author} {\bibfnamefont {M.}~\bibnamefont {Aguena}}, \bibinfo {author} {\bibfnamefont {S.}~\bibnamefont {Allam}}, \bibinfo {author} {\bibfnamefont {A.}~\bibnamefont {Amon}}, \bibinfo {author} {\bibfnamefont {F.}~\bibnamefont {Andrade-Oliveira}}, \bibinfo {author} {\bibfnamefont {J.}~\bibnamefont {Asorey}}, \bibinfo {author} {\bibfnamefont {S.}~\bibnamefont {Avila}}, \bibinfo {author} {\bibfnamefont {G.}~\bibnamefont {Bernstein}}, \bibinfo {author} {\bibfnamefont {E.}~\bibnamefont {Bertin}}, \bibinfo {author} {\bibfnamefont {A.}~\bibnamefont {Brandao-Souza}}, \emph {et~al.},\ }\bibfield  {title} {\bibinfo {title} {Dark energy survey year 3 results: A 2.7\% measurement of baryon acoustic oscillation distance scale at redshift 0.835},\ }\href@noop {} {\bibfield  {journal} {\bibinfo  {journal} {Physical Review D}\ }\textbf {\bibinfo {volume} {105}},\ \bibinfo {pages} {043512} (\bibinfo {year} {2022})}\BibitemShut
  {NoStop}%
\bibitem [{\citenamefont {VandenBerg}\ \emph {et~al.}(2014)\citenamefont {VandenBerg}, \citenamefont {Bond}, \citenamefont {Nelan}, \citenamefont {Nissen}, \citenamefont {Schaefer},\ and\ \citenamefont {Harmer}}]{vandenberg2014three}%
  \BibitemOpen
  \bibfield  {author} {\bibinfo {author} {\bibfnamefont {D.~A.}\ \bibnamefont {VandenBerg}}, \bibinfo {author} {\bibfnamefont {H.~E.}\ \bibnamefont {Bond}}, \bibinfo {author} {\bibfnamefont {E.~P.}\ \bibnamefont {Nelan}}, \bibinfo {author} {\bibfnamefont {P.}~\bibnamefont {Nissen}}, \bibinfo {author} {\bibfnamefont {G.~H.}\ \bibnamefont {Schaefer}},\ and\ \bibinfo {author} {\bibfnamefont {D.}~\bibnamefont {Harmer}},\ }\bibfield  {title} {\bibinfo {title} {Three ancient halo subgiants: precise parallaxes, compositions, ages, and implications for globular clusters},\ }\href@noop {} {\bibfield  {journal} {\bibinfo  {journal} {The Astrophysical Journal}\ }\textbf {\bibinfo {volume} {792}},\ \bibinfo {pages} {110} (\bibinfo {year} {2014})}\BibitemShut {NoStop}%
\bibitem [{\citenamefont {Catelan}(2017)}]{catelan2017ages}%
  \BibitemOpen
  \bibfield  {author} {\bibinfo {author} {\bibfnamefont {M.}~\bibnamefont {Catelan}},\ }\bibfield  {title} {\bibinfo {title} {The ages of (the oldest) stars},\ }\href@noop {} {\bibfield  {journal} {\bibinfo  {journal} {Proceedings of the International Astronomical Union}\ }\textbf {\bibinfo {volume} {13}},\ \bibinfo {pages} {11} (\bibinfo {year} {2017})}\BibitemShut {NoStop}%
\bibitem [{\citenamefont {Sahlholdt}\ \emph {et~al.}(2019)\citenamefont {Sahlholdt}, \citenamefont {Feltzing}, \citenamefont {Lindegren},\ and\ \citenamefont {Church}}]{sahlholdt2019benchmark}%
  \BibitemOpen
  \bibfield  {author} {\bibinfo {author} {\bibfnamefont {C.~L.}\ \bibnamefont {Sahlholdt}}, \bibinfo {author} {\bibfnamefont {S.}~\bibnamefont {Feltzing}}, \bibinfo {author} {\bibfnamefont {L.}~\bibnamefont {Lindegren}},\ and\ \bibinfo {author} {\bibfnamefont {R.~P.}\ \bibnamefont {Church}},\ }\bibfield  {title} {\bibinfo {title} {Benchmark ages for the gaia benchmark stars},\ }\href@noop {} {\bibfield  {journal} {\bibinfo  {journal} {Monthly Notices of the Royal Astronomical Society}\ }\textbf {\bibinfo {volume} {482}},\ \bibinfo {pages} {895} (\bibinfo {year} {2019})}\BibitemShut {NoStop}%
\bibitem [{\citenamefont {Ade}\ \emph {et~al.}(2019)\citenamefont {Ade}, \citenamefont {Aguirre}, \citenamefont {Ahmed}, \citenamefont {Aiola}, \citenamefont {Ali}, \citenamefont {Alonso}, \citenamefont {Alvarez}, \citenamefont {Arnold}, \citenamefont {Ashton}, \citenamefont {Austermann} \emph {et~al.}}]{ade2019simons}%
  \BibitemOpen
  \bibfield  {author} {\bibinfo {author} {\bibfnamefont {P.}~\bibnamefont {Ade}}, \bibinfo {author} {\bibfnamefont {J.}~\bibnamefont {Aguirre}}, \bibinfo {author} {\bibfnamefont {Z.}~\bibnamefont {Ahmed}}, \bibinfo {author} {\bibfnamefont {S.}~\bibnamefont {Aiola}}, \bibinfo {author} {\bibfnamefont {A.}~\bibnamefont {Ali}}, \bibinfo {author} {\bibfnamefont {D.}~\bibnamefont {Alonso}}, \bibinfo {author} {\bibfnamefont {M.~A.}\ \bibnamefont {Alvarez}}, \bibinfo {author} {\bibfnamefont {K.}~\bibnamefont {Arnold}}, \bibinfo {author} {\bibfnamefont {P.}~\bibnamefont {Ashton}}, \bibinfo {author} {\bibfnamefont {J.}~\bibnamefont {Austermann}}, \emph {et~al.},\ }\bibfield  {title} {\bibinfo {title} {The simons observatory: science goals and forecasts},\ }\href@noop {} {\bibfield  {journal} {\bibinfo  {journal} {Journal of Cosmology and Astroparticle Physics}\ }\textbf {\bibinfo {volume} {2019}}\bibinfo  {number} { (02)},\ \bibinfo {pages} {056}}\BibitemShut {NoStop}%
\bibitem [{\citenamefont {Liu}\ \emph {et~al.}(2022)\citenamefont {Liu}, \citenamefont {Sun}, \citenamefont {Han}, \citenamefont {Carron}, \citenamefont {Delabrouille}, \citenamefont {Li}, \citenamefont {Liu}, \citenamefont {Jin}, \citenamefont {Ghosh}, \citenamefont {Yue} \emph {et~al.}}]{liu2022forecasts}%
  \BibitemOpen
\bibfield  {number} {  }\bibfield  {author} {\bibinfo {author} {\bibfnamefont {J.}~\bibnamefont {Liu}}, \bibinfo {author} {\bibfnamefont {Z.}~\bibnamefont {Sun}}, \bibinfo {author} {\bibfnamefont {J.}~\bibnamefont {Han}}, \bibinfo {author} {\bibfnamefont {J.}~\bibnamefont {Carron}}, \bibinfo {author} {\bibfnamefont {J.}~\bibnamefont {Delabrouille}}, \bibinfo {author} {\bibfnamefont {S.}~\bibnamefont {Li}}, \bibinfo {author} {\bibfnamefont {Y.}~\bibnamefont {Liu}}, \bibinfo {author} {\bibfnamefont {J.}~\bibnamefont {Jin}}, \bibinfo {author} {\bibfnamefont {S.}~\bibnamefont {Ghosh}}, \bibinfo {author} {\bibfnamefont {B.}~\bibnamefont {Yue}}, \emph {et~al.},\ }\bibfield  {title} {\bibinfo {title} {Forecasts on cmb lensing observations with alicpt-1},\ }\href@noop {} {\bibfield  {journal} {\bibinfo  {journal} {Science China Physics, Mechanics \& Astronomy}\ }\textbf {\bibinfo {volume} {65}},\ \bibinfo {pages} {109511} (\bibinfo {year} {2022})}\BibitemShut {NoStop}%
\bibitem [{\citenamefont {Miao}\ \emph {et~al.}(2023)\citenamefont {Miao}, \citenamefont {Gong}, \citenamefont {Chen}, \citenamefont {Huang}, \citenamefont {Li},\ and\ \citenamefont {Zhan}}]{miao2023cosmological}%
  \BibitemOpen
  \bibfield  {author} {\bibinfo {author} {\bibfnamefont {H.}~\bibnamefont {Miao}}, \bibinfo {author} {\bibfnamefont {Y.}~\bibnamefont {Gong}}, \bibinfo {author} {\bibfnamefont {X.}~\bibnamefont {Chen}}, \bibinfo {author} {\bibfnamefont {Z.}~\bibnamefont {Huang}}, \bibinfo {author} {\bibfnamefont {X.-D.}\ \bibnamefont {Li}},\ and\ \bibinfo {author} {\bibfnamefont {H.}~\bibnamefont {Zhan}},\ }\bibfield  {title} {\bibinfo {title} {Cosmological constraint precision of photometric and spectroscopic multi-probe surveys of china space station telescope (csst)},\ }\href@noop {} {\bibfield  {journal} {\bibinfo  {journal} {Monthly Notices of the Royal Astronomical Society}\ }\textbf {\bibinfo {volume} {519}},\ \bibinfo {pages} {1132} (\bibinfo {year} {2023})}\BibitemShut {NoStop}%
\end{thebibliography}%
\end{document}